%% file: COVID_19_R10.tex
\documentclass{article}
\usepackage{amsfonts}
\usepackage{amsmath}
\usepackage{amssymb}
\usepackage{graphicx}
\usepackage{booktabs}
\providecommand{\U}[1]{\protect\rule{.1in}{.1in}}

\date{}
\begin{document}

\title{Dynamics of the COVID-19\\
Comparison between the Theoretical Predictions and the Real Data, and Predictions about Returning to Normal Life}
\author{Giorgio SONNINO and Pasquale NARDONE\\Facult{\' e} des Sciences\\
Universit\'{e} Libre de Bruxelles (ULB) \\ Boulevard de Triomphe, Campus Plain CP 231\\1050 Brussels, Belgium\\ Emails: gsonnino@ulb.ac.be and pnardon@ulb.ac.be\\
{}\\
Manuscript published in\\
\textit{ Annals of Clinical and Medical Case Reports}, {\bf 4} (9), 2020 \\ISSN 2639-8109}
\maketitle

\begin{abstract}
A new coronavirus disease, called COVID-19, appeared in the Chinese
region of Wuhan at the end of last year; since then the virus spread to other
countries, including most of Europe. We propose a differential equation
governing the evolution of the COVID-19. This dynamic equation also describes
the evolution of the number of infected people for 13 common respiratory
viruses (including the SARS-CoV-2). We validate our theoretical predictions with
experimental data for Italy, Belgium and Luxembourg, and compare them with the
predictions of the logistic model. We find that our predictions are in good
agreement with the real world since the beginning of the appearance of the
COVID-19; this is not the case for the logistic model that only applies to the
first days. The second part of the work is devoted to modelling the descending
phase, i.e. the decrease of the number of people tested positive for COVID-19.
Also in this case, we propose a new set of dynamic differential equations that
we solved numerically. We use our differential equations parametrised with
experimental data to make several predictions, such as the date when Italy,
Belgium, and Luxembourg will reach a peak number of SARS-CoV-2 infected people.
The descending curves provide valuable information such as the
duration of the COVID-19 epidemic in a given Country and therefore when it
will be possible to return to normal life. The study of the the dynamics of COVID-19 when the population have been subject to less restrictive measures is beyond the scope of this work and it will be matter of future works.
\vskip0.2cm
\noindent {\bf Key words}: Mathematical model; COVID-19; Pneumonia
\end{abstract}


\section{Introduction}\label{intro}
Viral infections usually affect the upper or lower respiratory tract. Although respiratory infections can be classified according to the causative agent (e.g. the flu), they are mostly clinically classified according to the type of syndrome (e.g., common cold, bronchiolitis, laryngo-tracheo-bronchitis acute, pneumonia).
Although pathogens typically cause characteristic clinical manifestations (e.g., rhinovirus causes the common cold, respiratory syncytial virus [RSV] usually causes bronchiolitis), they can all cause many of the most common respiratory syndromes. The severity of viral respiratory disease is highly variable; serious illness is more frequent in elderly patients and young children. Morbidity can either directly result from the infecting agent, or may be indirect. The latter case can be due to the exacerbation of an underlying cardiopulmonary disease, or a bacterial superinfection of the lung, paranasal sinuses, or middle ear. The main motivation of this work is to verify, by making theoretical predictions, that political decisions are truly effective to minimise the number of infected people in order to (i) not overload local health services (such as hospitals), and to (ii) gain time to allow research institutes to deliver vaccines or the anti-virals.

\input{Data_IT.tex}
\input{Data_IT_II.tex}
\input{Data_BE.tex}
\input{Data_BE_II.tex}
\begin{figure}
\hskip 2truecm
\includegraphics[width=6cm, height=8cm]{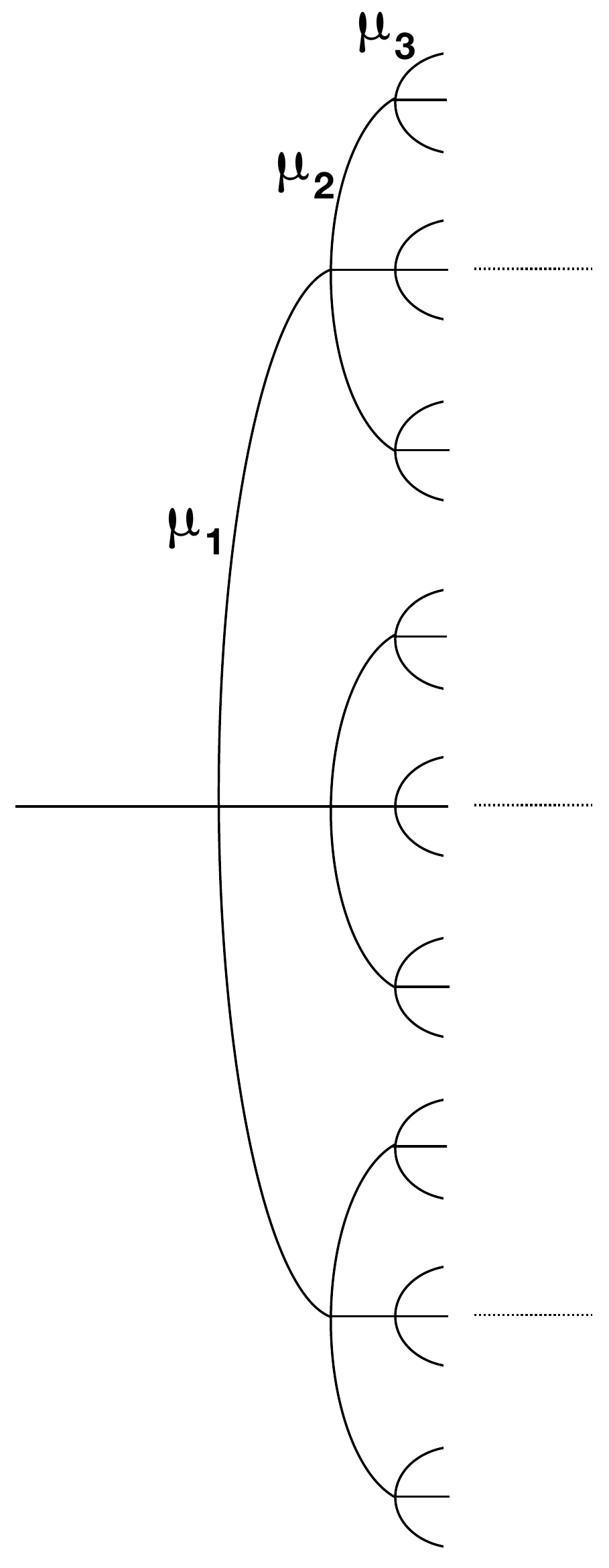}
\caption{
\textit{Schematic dynamics of respiratory virus in the absence of the lockdown
measures} In this graphics, for illustrative purpose only, we set $R_0=3$. However, for SARS-CoV-2, the value of $R_0$ is $2$ even at the beginning of the outbreak in China and Italy.  After a period of time $\mu_1$, an infected individual can infect $R_0$ other individuals. In turn, after a period $\mu_2$, each of these newly infected individuals can infect other $R_0$ people, and so on. After $n$ steps the elapsed time is $t=\sum_{i=1}^n\mu_i$.
}
\label{Fig.1}
\end{figure}

\noindent Tables~(\ref{A}) and~(\ref{B}) respectively provide the experimental data for Italy~\cite{sole24ore} and for Belgium~\cite{sciensano}, \cite{covid-19wiki}. They show the number of active people (i.e., people currently infected by SARS-CoV-2), the recovered people, and deaths for COVID-19.

\begin{figure*}[htb]
  \hfill
  \begin{minipage}[t]{.48\textwidth}
    \centering
    \includegraphics[width=5cm,height=5cm]{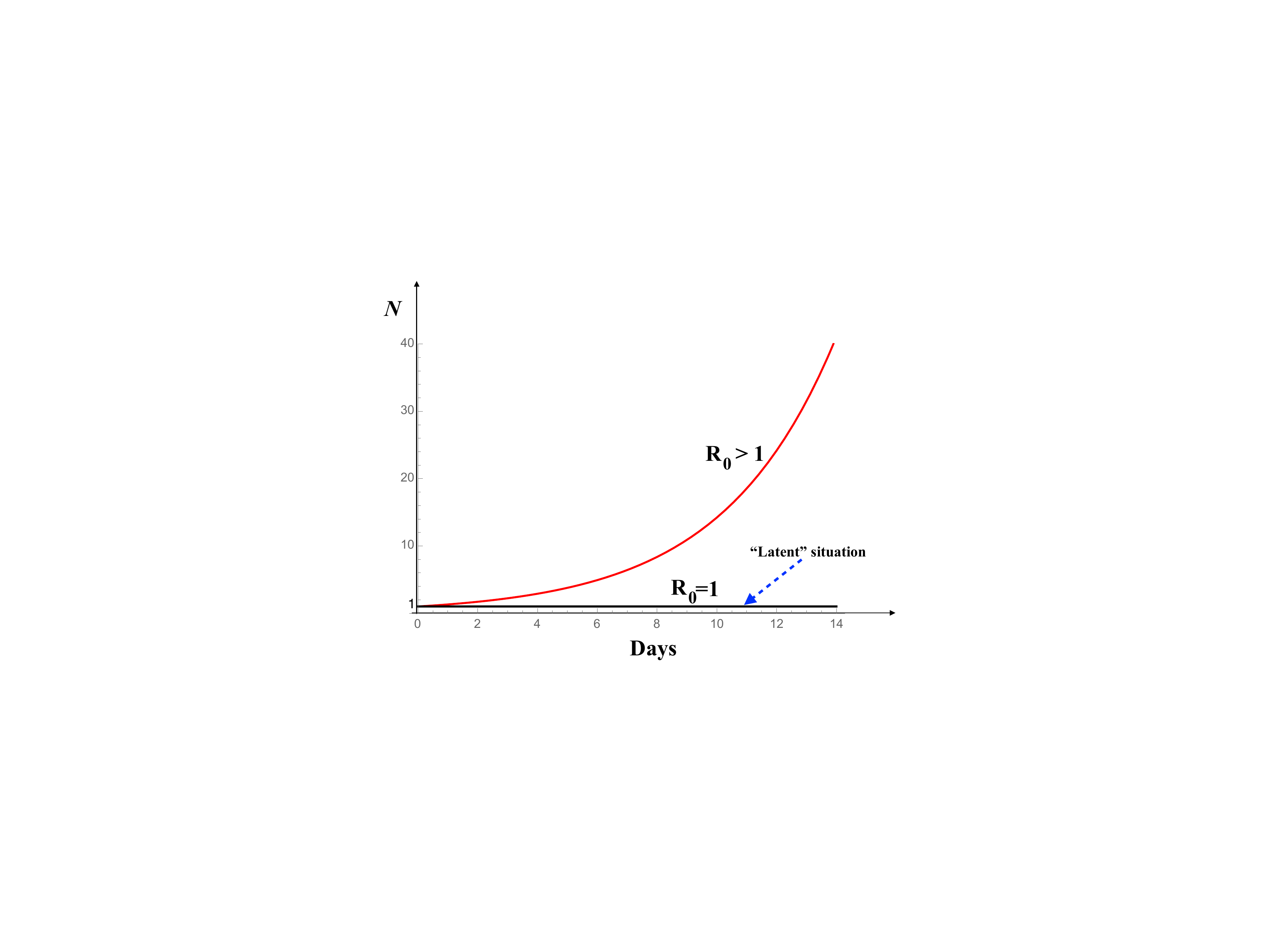}
    \caption{{\it Situation before the lockdown measures}. Number of infected people corresponding to the exponential law. The red line represents the case $R_0>1$, such as the situation before the adoption of lockdown measures. The black line corresponds to the case $R_0=1$, the {\it latent situation} in which the virus is substantially ineffective.}
    \label{Fig.2}
  \end{minipage}
  \hfill
    \begin{minipage}[t]{.48\textwidth}
      \centering
      \includegraphics[width=5cm,height=5cm]{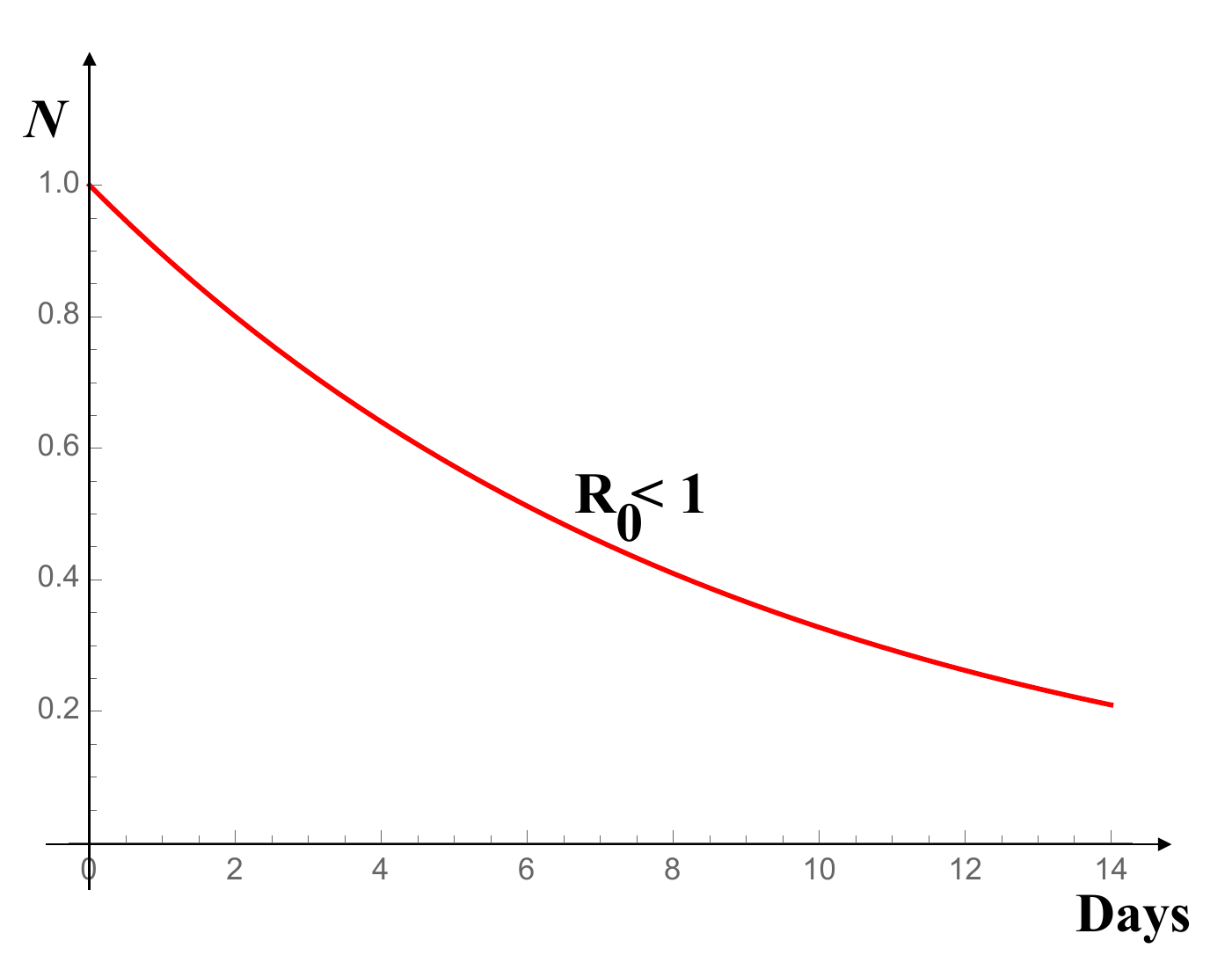}
      \caption{Number of infected people corresponding to the exponential law. The red line represents the case $R_0<1$. In this situation the number of infected people decreases exponentially and the virus disappears after a few weeks.}
      \label{Fig.3}
    \end{minipage}
  \hfill
\end{figure*}

\noindent We start our theoretical analysis by introducing the definition of the \textit {basic reproduction number of an infection} $R_0$, defined as \textit{the number of infected people derived from a first case in a population where all the others are susceptible}. So, it is not possible to modify $R_0$, in any case, but it is possible to get a different effective $R$\footnote{More rigourously, in epidemiology, \textit{the basic reproduction number of an infection, $R_0$, is the expected number of cases directly generated by one case in a population where all individuals are susceptible to infection in absence of any deliberate intervention in disease transmission} (see, for example, \cite{hystory}).}. This parameter is strictly linked to the {\it replication time of a virus}, indicated with $\mu_i$, defined as \textit{the time interval after which the number of infected people has increased by $R_0$ times}. Fig.~\ref{Fig.1} schematically represents the diffusion dynamics of the virus. By indicating with $N$ the number of infected people, after $n$ steps we get\footnote{In this Section we shall follow the definitions and the expressions reported in standard books or thesis dissertation such as, for example, \cite{murray} or \cite{lusuardi}.}:
\begin{equation}\label{I1}
N=R_0^n
\end{equation}
\noindent Of course, after $n$ steps, the elapsed time is $t = \sum_{i=1}^n\mu_i$ and, if there are $M$ outbreaks of infectious viruses, Eq.~(\ref{I1}) can be cast into the form\footnote{Actually, Eq.~(\ref{I2}) applies only if the $M$ outbreaks of the virus are exactly at the same conditions. In general, the correct expression reads $N=\sum_{i=1}^MR_0^{t/{\widetilde \mu}_i}$, with ${\widetilde \mu}_i$ indicating the replication time of the virus for the $i$-$th$ outbreak.}
\begin{equation}\label{I2}
N=M R_0^{t/\mu}
\end{equation}
\noindent with $\mu\equiv 1/n\sum_{i=1}^n\mu_i$. Note that the two parameters $R_0$ and $\mu$ are not independent (see, for example, \cite{anderson}, \cite{population} or \cite{gyorgy})\footnote{In ref.~\cite{anderson}, the doubling time is used to calculate $R_0$, by means of the equation $R_0=1+(\gamma+\rho)log(2)/\mu$ where $\gamma$ is the duration of the incubation period, $\rho$ is the duration of the symptomatic period, and $\mu$ is the doubling time (see \cite{anderson}). In this respect, we would also like to mention another excellent work recently produced by G. Steinbrecher~\cite{gyorgy}.}. It is more convenient to work in the Euler base $e$ rather than in base $R_0$; in the Euler base Eq.~(\ref{I2}) provides the {\it law of growth of a Malthusian population} \cite{murray}.
\begin{equation}\label{I3}
N=M\exp(t/\tau)\quad\textrm{ where}\ \ \tau=\frac{\mu}{\log(R_0)}
\end{equation}
\noindent In literature, $\tau$ is referred to as the \textit{characteristic time of the exponential trend}.  So, in the absence of containment measures the number of infected people follows the exponential law~(\ref{I3}). Let us now analyse Eq.~(\ref{I3}) in more dept. We have three possible scenarios:
\begin{enumerate}
 \item $R_0>1$ (as is the current world's situation). For Italy, for example, before the adoption of (severe) containment measures, the value of $\tau$ was about $\tau\sim 3.8$ days (and $\mu\sim 2.6$ days). In this case the number of the infected people increases exponentially.
 \item $R_0=1$ If the infection-capacity of the virus is of the type \textit{one-to-one} (i.e., a person infected by SARS-CoV-2 can in turn infects only another person), we get the stationary situation corresponding to $N = 1$. This situation is referred to as the {\it latent situation} i.e., the virus is still present but does not spread. In this limit case, the SARS-CoV-2 is substantially ineffective. Scenarios (1) and (2) are illustrated in Fig.~\ref{Fig.2}.
\item $0<R_0<1$. We may also imagine that the capacity of infection of SARS-CoV-2 is less than 1. This means that the virus is no longer able to be spread (e.g., thanks to protective measures, or to the production of vaccines and anti-virals, or because people who overcame the disease became immune. In this case, the value of $\tau$ is negative and the number of infected people decreases ever time. That is, the infection eventually disappears. The rate of decrease of the number of infected people depends on the value of $\tau$. This scenario is depicted in Fig.~\ref{Fig.3}.
\end{enumerate}

\subsection{Comparison with the Real Data for COVID-19 before the Lockdown Measures}
It is understood that the main objective of the lockdown measures established by most European governments and health organisations is to reduce the ability of a virus to spread. From a mathematical point of view, we would like to have $R_0=1$ (or, better, $R_0<1$), in Eq.~(\ref{I3}) instead of $R_0>1$. In practical terms, this means reducing the frequency of all involuntary contacts with a large number of people, reducing unnecessary movements to avoid encounters, and to prolong the closure of schools. Although these measures cannot prevent the spread of the infection in the long term, they can reduce the number of new infections daily. This has the benefit of leaving room for seriously-ill patients by avoiding to overload the healthcare system. We can easily realise what are the consequences if the lockdown measures are not set up. To make a comparison between the theoretical predictions and the experimental data in absence of lockdown measures, we have to consider the correct reference period. More specifically, we saw that the number of positive cases grows in the course of time by following the law~{(\ref{I3}). Hence, at the reference time $t_0$, the number of people infected by the virus is
\begin{equation}\label{I4}
N_0=M\exp(t_0/\tau)
\end{equation}
\noindent After a period of time, say $t$, Eq.~(\ref{I3}) reads
\begin{equation}\label{I5}
N=M\exp(t/\tau)
\end{equation}
\noindent Hence,
\begin{equation}\label{I6}
N=N_0\exp((t-t_0)/\tau)
\end{equation}
\noindent Eq.~({\ref{I6}) is the equation that we use for comparing the mathematical predictions with experimental data during the initial phase where the spread of SARS-CoV-2, causing the COVID-19, follows the exponential law, and $(t-t_0)$ is our \textit{reference period}. For the case of COVID-19 we get (see, for example, ~\cite{murray} or \cite{lusuardi}) 
\begin{itemize}
\item All infectious outbreaks are exactly at the same conditions. So, Eq.~(\ref{I2}) applies;
\item $R_0= 2$;
\item All the $\mu_i$ are equal with each other: $\mu_i=const=\mu$ (see also \cite{murray} and \cite{anderson}).
\end{itemize}
\noindent In this case, $\mu$ is referred to as the {\it doubling time}. So, the \textit {doubling time is the amount of time it takes for a given quantity to double in size or value at a constant growth rate} \cite{population}. If we do not apply the locking measures, the evolution in the course of time of the number of infected people is best approximated by an exponential curve with $R=2$, even though we have to stress that $R_0$ is only associated with the beginning of the epidemic and, with certain approximations, with the early stages, but not beyond. Fig.~\ref{Fig.6} and Fig.~\ref{Fig.7} respectively show the comparison between the theoretical predictions and the experimental data for Italy and Belgium before the lockdown measures. We get $\tau\simeq 3.8$ days and $\mu\simeq 2.6$ days for Italy, and $\tau\simeq 5.2$ days and $\mu\simeq 3.7$ days for Belgium. We conclude this Introduction by mentioning that there are several methods currently proposed in Literature to derive by mathematical models, the value of $R_0$. For example, in ref.~\cite{gyorgy}, we have a short numerical code, written in $R$-programming language for statistical computing and graphics, able to compute the estimated $R_0$ values for the following 17 infectious diseases: \textit{Chickenpox (varicella) (Transmission: Aerosol), Common cold (Transmission: Respiratory Droplets), COVID-19 (Transmission: Respiratory Droplets), Diphtheria (Transmission: Saliva), Ebola - 2014 Ebola outbreak (Transmission:: Body fluids, HIV/AIDS (Transmission: Body fluids), Influenza - 1918 pandemic strain (Transmission:: Respiratory Droplets), Influenza - 2009 pandemic strain (Transmission: Respiratory Droplets, Influenza - seasonal strains (Transmission: Respiratory Droplets), Measles (Transmission: Aerosol), MERS (Transmission: Respiratory Droplets), Mumps (Transmission: Respiratory Droplets), Pertussis (Transmission: Respiratory Droplets), Polio (Transmission: Fecal oral route), Rubella (Transmission: Respiratory Droplets), SARS (Transmission: Respiratory Droplets), Smallpox (Transmission: Respiratory Droplets)}. However, this task is particularly problematic if there are intermediate vectors between hosts, such as malaria. 
\begin{figure*}[htb]
  \hfill
  \begin{minipage}[t]{.45\textwidth}
    \centering
    \includegraphics[width=5cm,height=5cm]{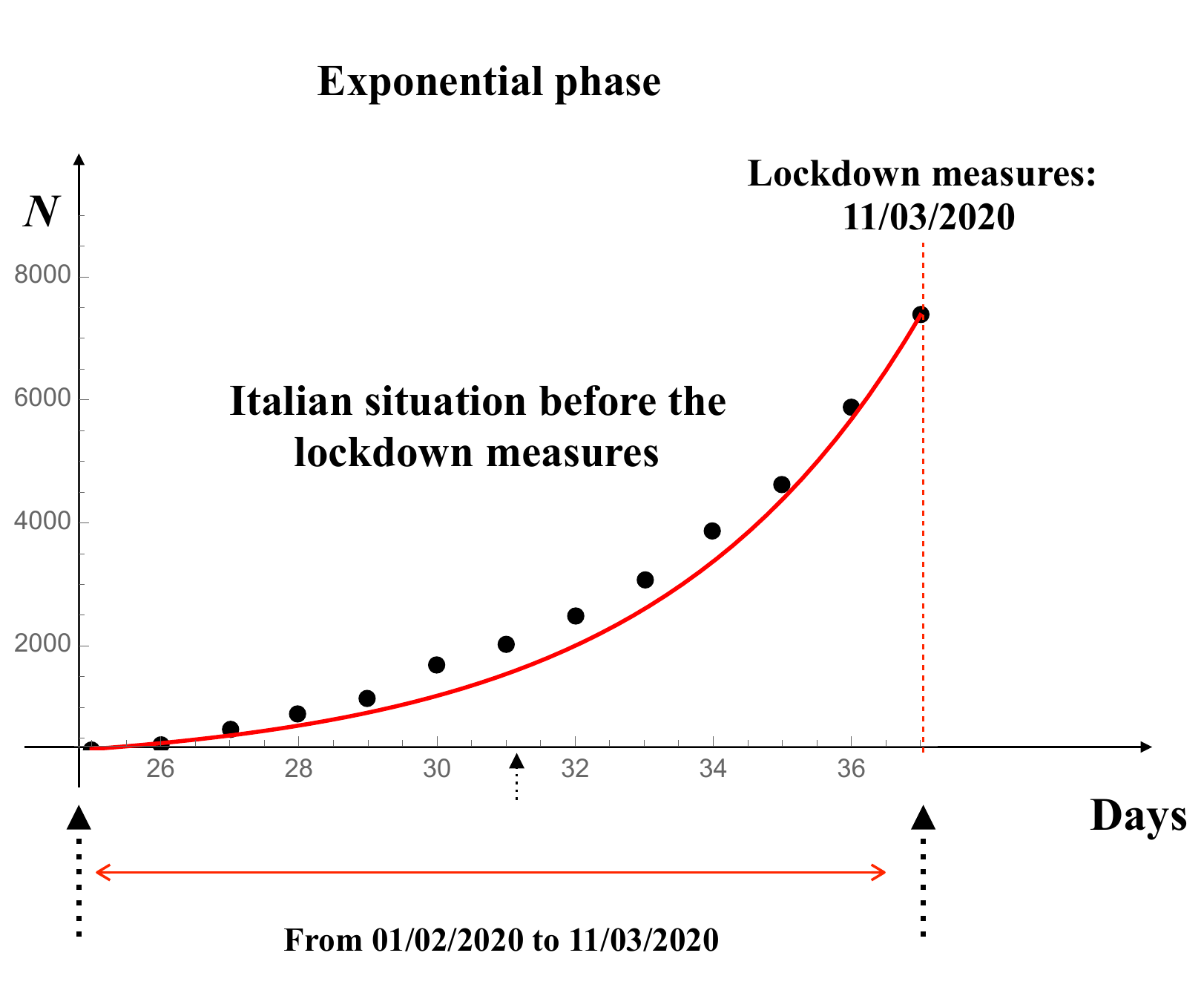}
    \caption{\textit{Number of infected people in Italy on the 10th of March 2020 (before the adoption of lockdown measures)}. The blue line corresponds to the theoretical predictions and the black dots correspond to experimental data. The values of the parameters $\tau_{IT}$ and $\mu_{IT}$ are $\tau_{IT}\simeq 3.8$ days and $\mu_{IT}\simeq 2.6$ days, respectively.}
    \label{Fig.6}
  \end{minipage}
  \hfill
  \begin{minipage}[t]{.45\textwidth}
    \centering
    \includegraphics[width=5cm,height=5cm]{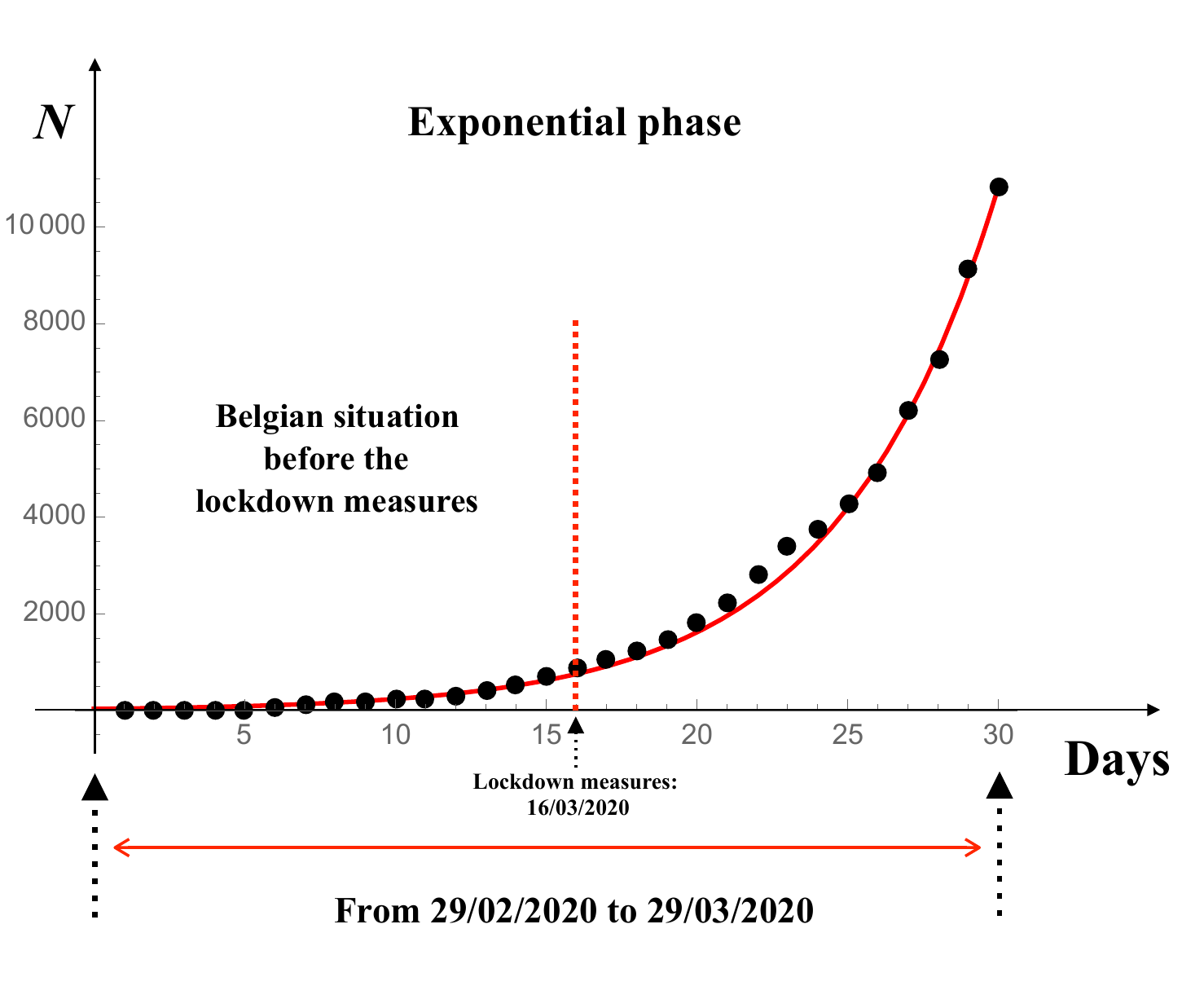}
    \caption{\textit{Exponential phase in Belgium}. The lockdown measure have been adopted on the 16the of March 2020 (however, initially not so strict as in Italy).The red line corresponds to the theoretical predictions and the black dots correspond to experimental data. The values of the parameters $\tau_{BE}$ and $\mu_{BE}$ are $\tau_{BE}\simeq 5.3$ days and $\mu_{BE}\simeq 3.7$ days, respectively.}
    \label{Fig.7}
  \end{minipage}
  \hfill
\end{figure*}

\noindent This manuscript is organised as follows. Section~(\ref{model}) determines the dynamic differential equation for the COVID-19; Section~(\ref{comparison}) compares the theoretical predictions and experimental data for Italy and Belgium. The differential equations providing the evolution of the decrease of the number of people tested positive for COVID-19 can be found in Section~(\ref{dp}); Section~(\ref{conclusions}) concludes. The comparison between  the theoretical predictions of our model and experimental data for Luxembourg, as well as the solution of the differential equations in the descending phase for Luxembourg are reported in Appendix. We may object that we are dealing data from countries which have passed the peak of infection, such as South Korea, Iceland or Austria etc. The situation in other Countries, which may have adopted other political decisions about the application of the lockdown measures, may be subject of future works. However, we would like to mention that several authors are currently applying our model to other Countries. In this regard, we have received their pre-prints such as the work cited in Ref.~\cite{mathus}. More specifically, we have received a message where our model has been used, with success, to analyse data from UK, USA, NY City, Spain, and Mumbai City. We stress the fact that this manuscript deals with the spread of SARS-CoV-2 until May 16, 2020, as the objective of this work is to study the effect of the strict lockdown measures. After May 2020, these measures have been modified by the various Governments, which have decided to introduce less heavy and much less restrictive lockdown measures\footnote{The study of the the dynamics of COVID-19 when the population has been subject to less restrictive measures is beyond the scope of this work and it will be matter of future works.}.

\section{Modelling the COVID-19 - Virus' growth}\label{model}
The objective of this section is to determine the coefficients of the evolutionary differential equation for the COVID-19 (see the forthcoming Eq.~(\ref{M4})). We also determine the generic analytical expression for the time-dependent number of infected people through fitting techniques validated by the $\chi^2$ tests. This expression is proposed after having previously analysed 12 respiratory infectious diseases caused by viruses~\cite{sonnino}, in addition of being solution to the Richard's differential equation.
\subsection{General background}
Letting $N$ represent population size and $t$ represent time, the Logistic model model is formalised by the differential equation below:
\begin{equation}\label{M1}
\frac{dN}{dt}=\alpha N\Bigl(1-\frac{N}{K}\Bigr)
\end{equation}
\noindent where $\alpha >0$ defines the \textit{grow rate} and $K>0$ is the \textit{carrying capacity}. In this equation, the early, unimpeded growth rate is modelled by the first term $+\alpha N$. The value of the rate $\alpha$ represents the proportional increase of the population $N$ in one unit of time. Later, \textit{if the system is closed} (i.e. the system is isolated and, hence, not in contact with a reservoir allowing the system to exchange individuals), as the population grows the modulus of the second term, $-\alpha N^2$/$K$, becomes almost as large as the first, until to saturating the exponential growth. This antagonistic effect is called the \textit{bottleneck}, and is modelled by the value of the parameter $K$. The competition diminishes the combined growth rate, until the value of $N$ ceases to grow (this is called {\textit{maturity of the population}). The solution of Eq.~(\ref{M1}) is
\begin{equation}\label{M1a}
N(t)=\frac{K}{1+B\exp{(- t/\tau)}}
\end{equation}
\noindent where $B>0$ is a constant related to the value of $N(0)$. It is more convenient to rewrite Eq.~(\ref {M1a}) in terms of the \textit{initial Logistic time} $t_ {0L}$ defined as
\begin{equation}\label{M1b}
t_{0L}=\tau\log B
\end{equation}
\noindent So, Eq.~(\ref{M1a}) may be cast into the form
\begin{equation}\label{M1c}
N(t)=\frac{K}{1+\exp{(- (t-t_{0L})/\tau)}}
\end{equation}
\noindent where $\tau$ is linked to the steepness of the curve. Since the environmental conditions influence the carrying capacity, as a consequence it can be time-varying, with $K(t)>0$, leading to the following mathematical model (see, for example, \cite{yukalov}):
\begin{equation}\label{M2}
\frac{dN}{dt}=\alpha N\Bigl(1-\frac{N}{K(t)}\Bigr)
\end{equation}
\noindent More generally, the growth modelling is well described by Richards' differential equation (RDE)~\cite{richards}
\begin{equation}\label{M3}
\frac{dN}{dt}=\alpha N\Bigl(1-\Bigl(\frac{N}{K(t)}\Bigr)^\nu\Bigr)
\end{equation}
\noindent where $\nu>0$ affects near which asymptote maximum growth occurs. The phenomenological logistic function is used to model the evolution of the COVID-19 pandemic in different Countries. The logistic model is mainly used in epidemiology and provides insights into the transmission dynamics of the virus. We note, however, to evaluate the dynamics of transmission of SARS-CoV-2, more refined models are needed, which take into account specific measures adopted in each Country \cite{ruffini}. So, let us suppose that the Government decides to adopt the lockdown measures. After the application of the lockdown measures the equation may be revised to be
\begin{equation}\label{M4}
\frac{dN}{dt}=\alpha N\Bigl(1-\Bigl(\frac{N}{K(t)}\Bigr)^\nu\Bigr)-c(t) N
\end{equation}
\noindent where $c (t)$ takes into account the degree of effectiveness of the lockdown measures.

\subsection {Determination of the Carrying Capacity and the Lockdown Coefficient for the COVID-19}
According to ref.~\cite{ronchetti}\footnote{Prof. Roberto Ronchetti is currently working at the Pediatric Clinic of \textit{La Sapienza} of the University of Rome, at the \textit{Policlinico Umberto I} and at the \textit{S. Andrea 24 MarchHospital} where he helped to found, dealt with childhood respiratory diseases, and studied bronchiolitis in particular. In these days he has studied (with his collaborators) the data available on SARS-CoV-2 in China, in South Korea and now in Italy.} \textit{Respiratory viruses remain quiet for months, inactive but viable, within living cells. Then suddenly they activate and become virulent as they say, the infectious capacity grows to a maximum, after which it decreases. The time duration is about of 2 or 3 months. So we can expect that the epidemic will soon die out in Italy too. So, there is no valid reason to think that this coronavirus behaves differently from others}~\cite{ronchetti}. The present work starts from the following hypothesis: \textit{the SARS-CoV-2 behaves like the other viruses that cause respiratory diseases}. As a consequence, for the COVID-19 case, functions $K(t)$ and $c(t)$ are determined by performing several \textit{fittings} on the growth rate-trends of infection capacity of the viruses that mainly affect the respiratory system. More specifically, we considered the following 13 different diseases caused by 12 different viruses: \textit{Whooping Cough (Pertussis)}, \textit{Swine Flu (H1N1)}, \textit{Bird Flu (Avian Flu H5N1)}, \textit{Enterovirus}, \textit{Flu in Children}, \textit{Flu in Adults}, \textit{Bacterial Pneumonia}, \textit{Viral Pneumonia}, \textit{Bronchitis}, \textit{Common Cold (Head Cold)}, \textit{Severe acute respiratory syndrome (SARS)}, and \textit{MERS (Middle East Respiratory Syndrome)}.
In all the examined cases, we took into account the fact that the therapy-induced death rate is greater than the baseline proliferation rate, then there is the eradication of the disease. In other words, for the above-mentioned cases the function $c(t)$ in Eq.~(\ref{M4}) represents the therapy-induced death rate~\cite{hilbe}-\cite{galvani}. Of course, this is an oversimplified model of both the growth and the therapy (e.g., it does not take into account the phenomenon of clonal resistance). We empirically noticed (according to the $\chi^2$ test) that all these viruses have in common the same growth rate-trend of infected people (of course, each of these behaviours have their own growth rate parameters). In particular, we get that the trends of the total number of infected people by respiratory viruses (indicated with $N$), subject to the therapy-induced death rate, versus time satisfy the following O.D.E.~\cite{sonnino}
\begin{equation}\label{MN9}
\frac{dN}{d{\widehat t}}=\alpha N\Bigl(1-\frac{N}{K_N}\Bigr)-\Bigl(\frac{\alpha {\widehat t}^2- 1}{{\widehat t}}\Bigr)N\qquad{\rm with}\quad {\widehat t}> 1/\alpha^{1/2}
\end{equation} 
\noindent where we have introduced the dimensionless time ${\widehat t}\equiv t/t_0$. The coefficient
\begin{equation}\label{MN10}
c(t)\equiv\Bigl(\frac{\alpha {\widehat t}^2- 1}{{\widehat t}}\Bigr)\qquad({\rm with}\quad {\widehat t}> 1/\alpha^{1/2})
\end{equation} 
\noindent is referred to as the \textit{average therapy-induced death rate}. In our case the term $-c(t)N$ in the dynamic equation represents the lockdown measures. The lockdown is mainly based on the isolation of the susceptible individuals, eventually with the removal of infected people by hospitalisation\footnote{It is worth mentioning that initially England did not adopt any lockdown measures believing that the British system be a closed system. Basically, it was believed that the system be governed by a simply logistic equation with a carrying capacity decreasing in time. However, contrary to the expectations, in England the carrying capacity did not decrease in time. This induced the British government to adopt the lockdown measures.}. In the idealised case, for $\alpha {\widehat t}^2>1$, $c(t)$ may be modelled as a linear function of ${\widehat t}$, by getting
\begin{equation}\label{MN11}
c(t)\simeq\alpha {\widehat t}
\end{equation} 
\noindent As for the epidemiological explanations relating to the various modelling of $ c (t) $ (constant, linear in time etc.), we refer the reader to the well-known and extensive literature on the subject (see, for example, Ref~\cite{sun} or to the references cited in \cite{wiki}). Here, we limit ourselves to give a very intuitive explanation on the physical meaning of this contribution. Immediately after the lockdown measures have been adopted, i.e. during the very first initial phase, we expect that $c({\hat t})$ is practically constant in time, as these measures have not yet been able to act effectively. However, after a short period of time, the positive effects of the lockdown measures become increasingly efficient and it is intuitive to expect a linear growth of $c ({\widehat t})$ in time. More specifically, we expect that, after a short transition period the coefficient $c (t)$ starts to grow linearly in time. Successively, at the leading order, $c ({\widehat t})$ will be equal in magnitude to the coefficient of the linear term (in order to balance the growth rate induced by the linear term). This because the lockdown measures will be able to satuarte the exponential growth. Briefly, we expect that the O.D.E. governing the dynamics of the SARS-CoV-2 is of the form~(\ref{MN9}) where $c({\widehat t})\simeq \alpha {\widehat t}$ for ${\widehat t} > 1/\alpha^{1/2}$. Indeed, for values of time ${\widehat t}\simeq 1$, the lockdown term in Eq.~(\ref{MN9}) is able to balance the exponential grow, which is in agreement with our intuitive expectations.

\noindent From Eq.~(\ref{MN9}) we get that the time derivative of $N$ vanishes for $N=N_s$, with $N_s$ satisfying the equation
\begin{equation}\label{MN11a}
\frac{N_s}{K_N}=\frac{\alpha {\hat t}-\alpha {\hat t}^2+1}{\alpha{\hat t}}>0
\end{equation} 
\noindent By taking into account the inequality reported in Eq.~(\ref{MN10}), we get that the O.D.E.~(\ref{MN9}) is valid in the range
\begin{equation}\label{MN11b}
\frac{1}{\alpha^{1/2}}<{\hat t}<\frac{(1+\alpha/4)^{1/2}}{\alpha^{1/2}} +\frac{1}{2}
\end{equation} 
\noindent Parameters $K_N$ and $\alpha$ depend on the virus in question and on the external conditions (e.g. in our case, the lockdown measures) to which the population is subject. In Eq.~(\ref{MN9}), the term $-N^ 2/K_N$ is the term that tends to saturate the exponential growth. $K_N$ is constant (or decreases) in the course of time since the non-linear contribution becomes more and more important until saturating the exponential growth. In our model, the carrying capacity is kept constant. 

\noindent For large values of the carrying capacity, the solutions of Eq.~(\ref{MN9}) reach the plateau at the time ${\hat t}_{Max}$ given, approximatively, by the expression
\begin{equation}\label{M10}
{\hat t}_{Max}\equiv\frac{t_{Max}}{t_0}\simeq \frac{1}{\alpha^{1/2}}+\frac{1}{2}
\end{equation}
\noindent Notice that $\alpha$ is linked to $\mu$. Indeed, as shown in Section~\ref{intro}, during the exponential period the COVID-19 grows according to the law (see Eq.~(\ref{I6})):
\begin{equation}\label{M9a}
\frac{dN}{d{\hat t}}\simeq {\widehat \tau}^{-1}N\quad {\rm where} \ \ {\widehat\tau}\equiv\frac{\tau}{t_0}
\end{equation}
\noindent Hence, we get
\begin{equation}\label{M9b}
\alpha\simeq\frac{1}{{\widehat\tau}}=\frac{\log(R_0)}{2{\widehat {\mu}}}\quad{\rm where}\quad {\widehat {\mu}}\equiv \frac{\mu}{t_0}
\end{equation}
\noindent We conclude this Section by mentioning that we may easily check that, for systems having a large carrying capacity, the solution of Eq.~(\ref{MN9}) is well approximated by the expression
\begin{equation}\label{M5}
N\simeq At\exp(-(t-t_0)^2/\sigma)\quad{\rm with}\quad \sigma=2t_0^2/\alpha
\end{equation}
\noindent The values of parameters $A$, $t_0$ and $\sigma$ depend on the virus in question. It is convenient to re-write Eq.~(\ref{M5}) in dimensionless form
\begin{equation}\label{M5bis}
N\simeq {\widehat A}{\widehat t}\exp(-({\widehat t}-1)^2/{\widehat \sigma})\quad{\rm where} \ \ {\widehat A}\equiv At_0\ ; \ {\widehat \sigma}\equiv \frac{\sigma}{t_0^2}
\end{equation}
\noindent To sum up, according to our model \textit{for COVID-19, the ascending behaviour of the total cases (i.e., the number of of positive cases plus the cumulative number of recovered people plus the cumulative number of deaths) is given by the solution of Eq.}~(\ref{MN9}) \textit{for} $1/\alpha^{1/2}\leq {\hat t}\leq {\hat t}_{Max}$. 

\noindent Notice that the determination of the O.D.E.~(\ref{MN9}) is of a fundamental importance if we wish also to describe the stochastic process (and the associated Fokker-Planck equation) where a white noise is added to this O.D.E. According to the literature nomenclature, we refer to the differential equation~(\ref{MN9}) as {\it COVID-19 dynamic model}\footnote{Viral dynamics is a field of applied mathematics concerned with describing the progression of viral infections within a host organism (see, for example, \cite{novak}.)}.

\subsection{Comparison between the Theoretical Predictions and Experimental Data}\label{comparison}
For Italy and Belgium one observes two distinct phases related to the dynamics of the COVID-19, which we classify as \textit{before the adoption of the lockdown measures} and \textit{after some days after the adoption of the lockdown measures}. The question therefore naturally arises, of whether these two types of regime are separated by a well-defined transition. We shall see that this is indeed the case. We may identify three different periods, which may be classified as follows:
\vskip0.1cm
\begin{figure*}[t]
  \hfill
    \begin{minipage}[t]{.45\textwidth}
    \begin{center}
      \includegraphics[width=5cm,height=5cm]{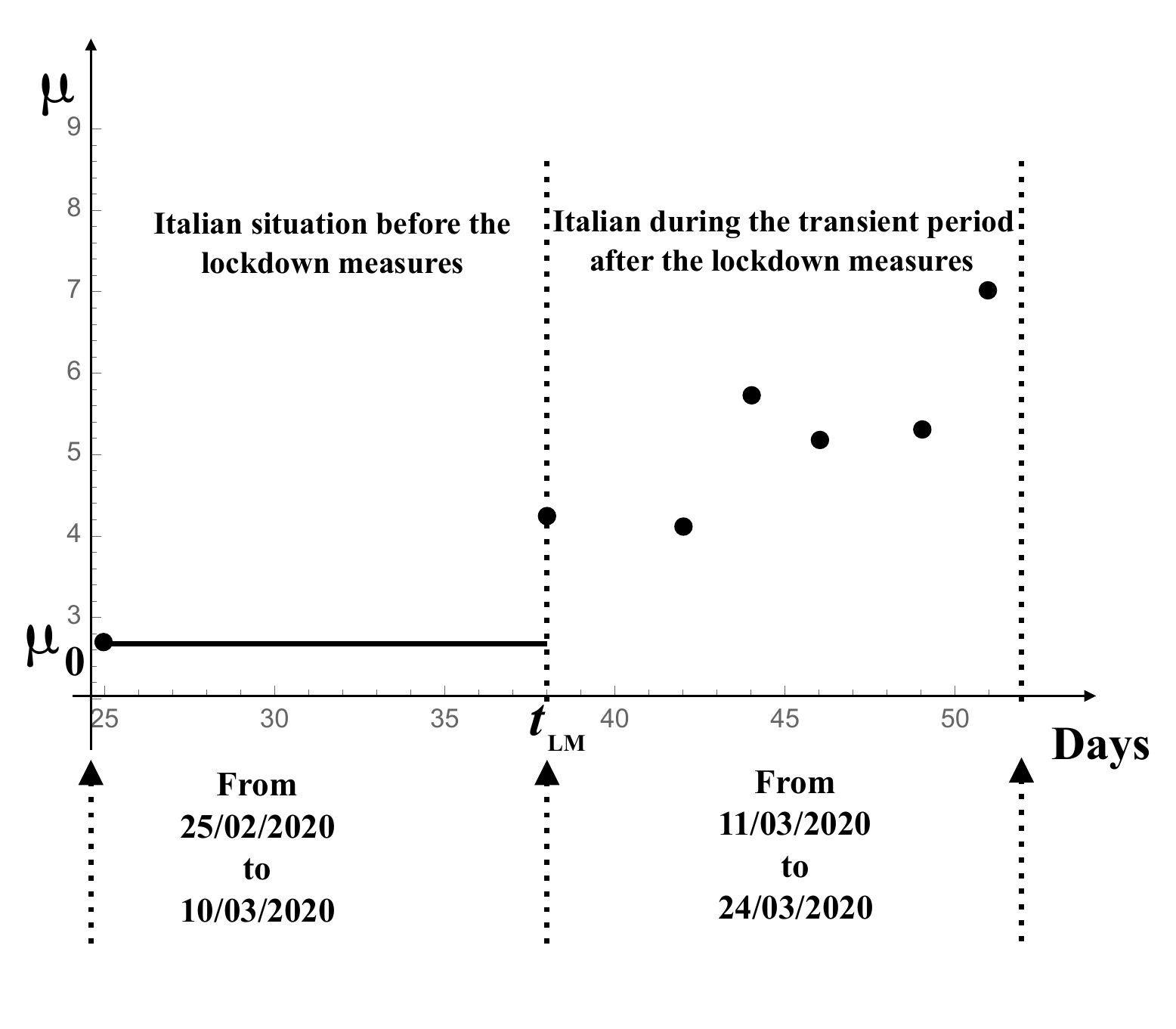}
      \caption{\textit{Italian transient period (from the 10th of March 2020 to the 24th of March 2020)}. During this period, the doubling time $\mu$ oscillates over time. $\mu_0$ indicates the (constant) doubling time during the exponential period (for Italy $\mu_0\simeq 2.6$ days).}
      \label{Fig.8}
    \end{center}
  \end{minipage}
  \hfill
  \begin{minipage}[t]{.45\textwidth}
    \begin{center}
      \includegraphics[width=5cm,height=5cm]{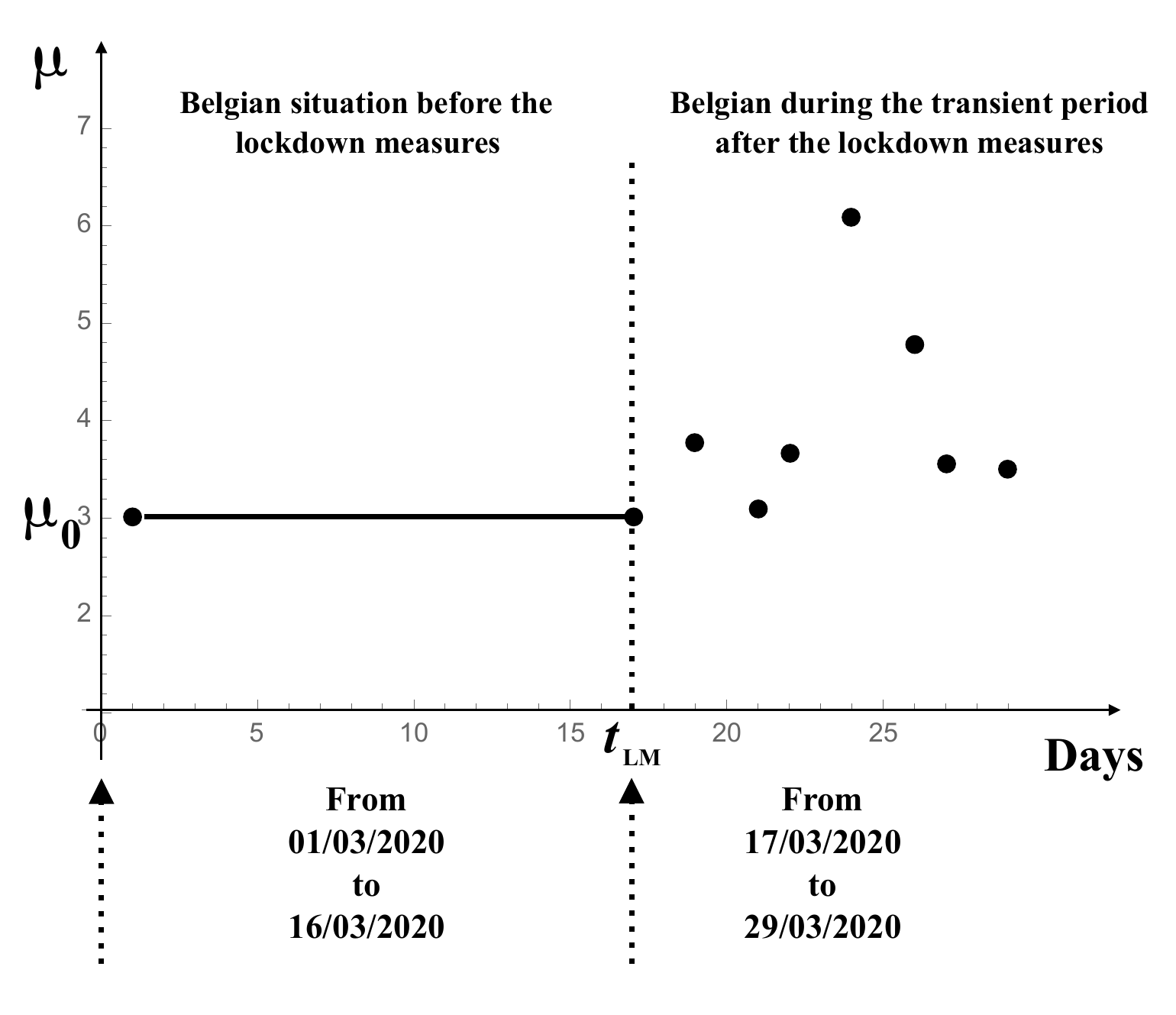}
      \caption{\textit{Belgian transient period (from the 17th of March 2020 to the 29th of March 2020)}. During this period, the doubling time $\mu$ oscillates over time. $\mu_0$ indicates the (constant) doubling time during the exponential period (for Belgium $\mu_0\simeq 3.7$ days).}
      \label{Fig.9}
    \end{center}
  \end{minipage}
  \hfill
\end{figure*}
\noindent
\begin{enumerate}

\item The \textit{exponential period}. As seen in Section~\ref{intro}, before the adoption of lockdown measures, the \textit{exponential trend} is the intrinsic behaviour of the grow rate of the COVID-19. In this period the doubling time $\mu$ is a \textit{constant parameter} versus time.\noindent
\item The \textit{transient period}. The transient period \textit{starts} after having applied the severe lockdown measures. In this period we observe a sort of oscillations (or fluctuations) of $\mu$ versus time. In this case the time variation of $\mu(t)$ reflects the behaviour of the \textit{time effective reproduction number}, $R(t)$, defined as the \textit{number of cases generated in the current state of a population, which does not have to be the uninfected state}. Fig.~\ref{Fig.8} and Fig.~\ref{Fig.9} show the behaviour of the parameter $\mu$ versus time for Italy and Belgium, respectively. The transient period \textit{ends} when the last step of the exponential trend fits real data as good as the linear trend\footnote{A numerical condition may be established by using the $\chi^2$ test: the fittings of the two trends are considered both \textit{good} if, for example, for both trends, the $\chi^2$-tests get values $\ge 0.9$.}.

\item The \textit{bell-shaped period} (or the \textit{post-transient period}). In the bell-shaped period parameter $\mu$ is a (typical) function of time obtained by using Eq.~(\ref{MN9}). Several theoretical models can be used to study the post-transient period (e.g., by using Gompertz's law~\cite{gompertz}). Here, we use two mathematical models: the solution of the differential equation~(\ref{MN9}) and the \textit{logistic model} (see, for example, Ref.~\cite{hilbe}), and we compare these two theoretical models with real data for Italy and Belgium.
\end{enumerate}
Figures~(\ref{Fig.10}), (\ref{Fig.11}), and (\ref{Fig.18}) (see Appendix) compare the predictions of our model (blue lines) against the logistic model (red lines) for Italy, Belgium, and Luxembourg, respectively. Notice that the number of free parameters of these two models are exactly the same, since $\tau$ and $\alpha$ cannot be chosen arbitrarily. More specifically,
\begin{description}
\item{{\bf a)}} The logistic model possesses two free parameters : $K$ and $t_{0L}$. Notice that parameter $\tau$ is not free since it is linked to the doubling time $\mu$;
\item{{\bf b)}} Also our model possesses two free parameter: $K_N$ and $t_0$. Notice that parameter $\alpha$ is linked to the doubling time $\mu$ (see Eqs~(\ref{MN9}) and (\ref{M9a})).
\end{description}
\begin{figure}[th!]
\hskip 2.5truecm
\includegraphics[width=7cm, height=7cm]{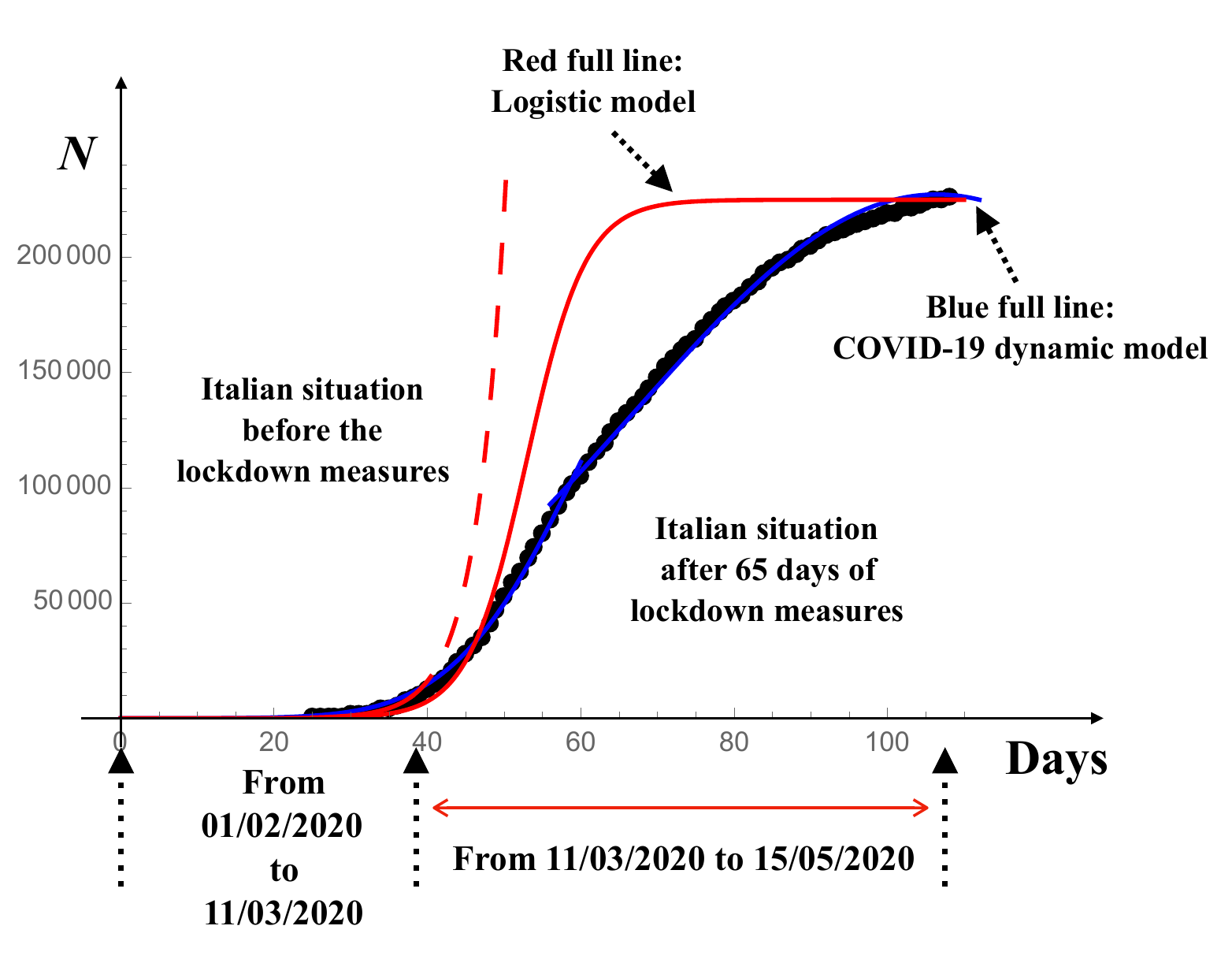}
\caption{\textit{Situation in Italy on 15 May 2020---before, and 65 days after, the adoption of lockdown measures}. The black dots correspond to experimental data. The red dotted line corresponds to the situation in Italy before the adoption of the lockdown measures. The blue and the red solid lines correspond to the theoretical predictions for Italy according to the solution of Eq.~(\ref{MN9}) and the logistic model, respectively. Solution of Eq.~(\ref{MN9}) fits well all the experimental data from the initial days (i.e., from the 1st of February 2020), while the logistic model applies only to the first days. The values of the parameters of Eq.~(\ref{MN9}) and the logistic function~(\ref{M1c}) are: $\tau_{IT}\simeq 3.8$ days ($\mu_{IT}=2.6$ days), $K_N^{IT}\simeq 355250$, and $t_{0IT}\simeq 72.5$ days for Eq.~(\ref{MN9}), and $\tau_{IT}\simeq 3.8$ days ($\mu_{IT}=2.6$ days), $K_{IT}=225000$, $t_{0LIT}=53$ days for the Logistic function.}
  \label{Fig.10}
\end{figure}
\begin{figure}[th!]
\hskip 2.5truecm
\includegraphics[width=7cm, height=7cm]{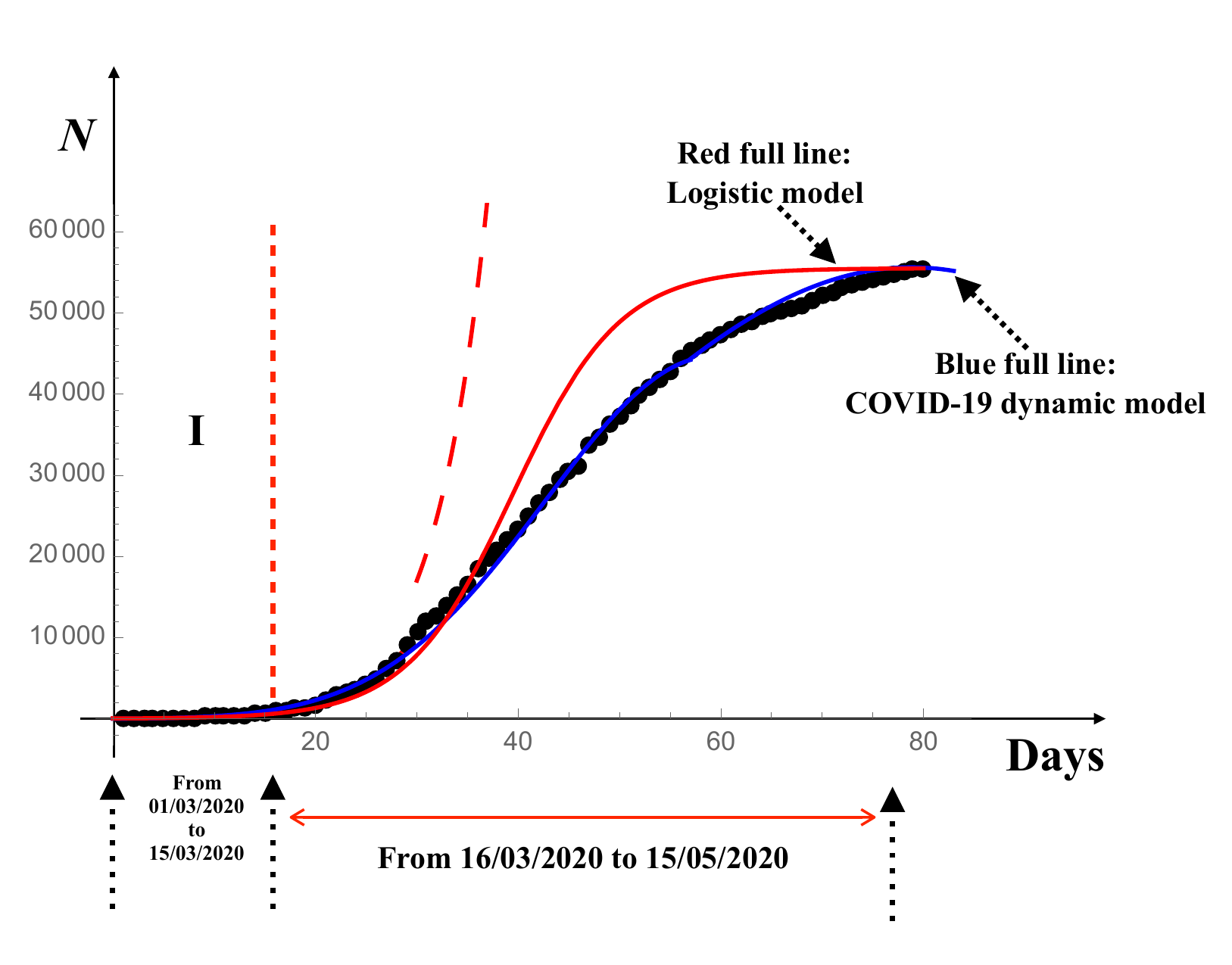}
\caption{\textit{Situation in Belgium on 15 May 2020---before, and 60 days after, the adoption of lockdown measures}. The black dots correspond to real data. The blue dotted line corresponds to the situation in Belgium before the adoption of the lockdown measures. The blue and the red solid lines correspond to the theoretical predictions for Belgium according to the solution of Eq.~(\ref{MN9}) and the logistic model, respectively. Solution of Eq.~(\ref{MN9}) fits well all the experimental data from the initial days (i.e., from the 29th of February 2020), while the logistic model applies only to the first data. The values of the parameters of Eq.~(\ref{MN9}) and the logistic function~(\ref{M1c}) are: $\tau_{BE}\simeq 5.3$ days ($\mu_{BE}=3.7$ days), $K_N^{BE}\simeq 42626$, and $t_{0BE}\simeq 53.4$ days for Eq.~(\ref{MN9}), and $\tau_{BE}\simeq 5.3$ days ($\mu_{BE}=3.7$ days), $K_{BE}=111000$, $t_{0LBE}=39.5$ days for the Logistic function, respectively. The zone $I$ corresponds to the period before the adoption of the lockdown measures.}
\label{Fig.11}
\end{figure}
\noindent Fig.~\ref{Fig.10} and Fig.~\ref{Fig.11} compare the theoretical predictions, with the experimental data for Italy and Belgium updated to the 15th of May 2020. The values of the parameters $\tau$, $K_N$, and $t_0$ for Eq.~(\ref{MN9}) and the parameters $\tau$, $t_0$ and $K$ for the logistic function are reported in the figure captions. As we can see, for both Countries Eq.~(\ref{MN9}) fits well all the real data from the initial days, while the logistic model applies only to the first data. The curves reach the plateau at the time $t_{Max}$ given by Eq.~(\ref{M10}). By inserting the values of the parameters, we get
\begin{equation}\label{c2}
t_{MaxIT}\simeq 80\ \textrm{ days}\quad\textrm{ and}\quad t_{MaxBE}\simeq 60\ \textrm{ days}
\end{equation}
\noindent corresponding to $t_{MaxIT}\simeq 21\ \textrm{ April}\ 2020$ and $t_{MaxBE}\simeq 2\ \textrm{ May}\ 2020$ for Italy and Belgium, respectively.

\section{Modelling the COVID-19 - The Descending Phase}\label{dp}
Here, for the descending phase is intended the phase where the number of the positive cases starts to decrease\footnote{We define the number of \textit{positive people} as the \textit{number of people tested positive for COVID-19}, hence, by excluding the number of the deceased people and the number of people who recovered.}. So, our model cannot be used for describing also the descending phase since $N_t$ is the number of the total cases and, during the descending phase, $N_t$ tends to reach the \textit{plateau}. The objective of this section is to determine the trend of the curve of \textit{positive people} during the descending phase. This task is accomplished by establishing the
appropriate equations for the {\it recovered people} and the {\it deceased
people} for COVID-19. During the descent phase the number of active people over
time must satisfy a conservation equation. This allows determining the time-
evolution for the positive people. In the sequel, we denote with $r_t$, $d_t$, and $n_t$ the \textit{number of people released from the hospital at the time} $t$, the \textit{total deaths}, and the number of \textit{positive individuals} at time $t$, respectively\footnote{We draw the attention of the reader that in this manuscript $N_t$ (capital letter) represents the \textit{number of the total cases at time} $t$, whereas $n_t$ (small letter) refers to the \textit{number of the positive individuals at time} $t$.} 

\subsection{Number of the Recovered People}
\noindent We start with the recovered people previously hospitalised. Let us suppose that a hospital has 50 patients in intensive therapy, corresponding to its maximum availability capacity. If the hospital is unable to heal any patient, the growth rate of healed people is necessarily equal to zero. On the other hand, if the hospital is able to heal a certain number of people, the places previously occupied by the sick people will free and other patients affected by COVID-19 will be able to be hospitalised. In the latter case, the growth rate of the healed people will rise thanks to the fact that the hospital is able to heal more and more patients. This initial phase may be modelled by introducing into the dynamic equation the term $+\gamma r(t)$, with $r(t)$ indicating the \textit{number of the recovered people at the time $t$, previously hospitalised} 
\begin{equation}\label{r1}
\frac{dr_t}{dt} = \zeta r_t
\end{equation}
\noindent However, it is reasonable to suppose that $\zeta$ is constant for low values of $r_t$, whereas, when $r_t$ takes more and more large values, $\zeta$ is proportional to $I_r$, with $I_r$ denoting the number of the infected people \textit{entering in the hospital} (and not the total number of the infected people, which is indicated with $n_t$). Hence,
\begin{equation}\label{r2}
\zeta=\alpha_r-\beta_rI_r>0
\end{equation}
\noindent The sign minus in Eq.~(\ref{r2}) is due to the fact that the recovered people will continue to grow linearly until when it reaches a maximum limit i.e. until when the hospital is no longer able to accept other sick people; this causes a reduction of people who recover. The competition between these two effects diminishes the combined growth rate. Hence,
\begin{equation}\label{r3}
I_r<\frac{\alpha_r}{\beta_r}
\end{equation}
\noindent where $\beta_r$ is linked to the hospital's capacity to accept sick people\footnote{Indeed, $\beta_r$ is proportional to the inverse of the hospital capacity.}. To sum up, 
\begin{equation}\label{r4}
\frac{dr_t}{dt} =\alpha_r r(t)-\beta_rI_r(t-\theta)r(t)\qquad{\rm where}\ \ I_r(t-\theta)=r(t)+D_r(t+\theta_1)
\end{equation}
\noindent $r(t)$ is the number of the recovered people, previously hospitalised, at the time $t$ who have been infected, in average, at the time $t-\theta$ and $D_r(t+\theta_1)$ denotes the number of deceased people at the hospital at the time $t+\theta_1$ who have been infected, in average, at the time $t-\theta$ (in general, $\theta_1\neq 0$). 

\noindent Clearly, the \textit{number of the recovered people, previously hospitalised, at the step} $n$ (i.e. $r_n$), is linked to the \textit{{\bf total} number of the recovered people previously hospitalised} at the step $n$ (denoted by $h_n$) by the relation
\begin{equation}\label{dp1a}
r_n=h_n-h_{n-1}\quad{\rm or}\quad h_t=\sum_{n=1}^{n=t/\Delta t} r_n\quad ({\rm with}\ \  \Delta t\simeq 1\ {\rm day})
\end{equation}
\noindent where we have set $h_0=0$. 
\subsubsection{Approximated O.D.E. for the Recovered People Previously Hospitalised}

\noindent We assume that \textit{all the infected people entering in the hospitals will heal}. So 
\begin{equation}\label{dp1abis}
D_r(t+\theta_1)\approx 0\quad{\rm hence}\ \ I_r(t-\theta)\simeq r(t)
\end{equation}
\noindent The final O.D.E. for the recovered people reads then
\begin{equation}\label{dp1}
\frac{dr_t}{dt}\simeq\alpha_r \Big(1-\frac{1}{K_r}r_t\Bigr)r_t
\end{equation}
\noindent where $K_r$ is the hospital's capacity, which we assume to be a time-independent parameter. Notoce that, under this approximation, the equation for the number of recovered people is \textit{in itself and independent of the equations for the other variables} (i.e. for $n_t$ and $d_t$)\footnote{Eq.~(\ref{dp1}) models a hospital's ability to heal people and, by no means, it must be linked to the number of people tested positive for COVID-19 or to the mortality rate caused by the SARS-CoV-2.}. 

\subsubsection{O.D.E. for the Total Recovered People}

\noindent At the first approximation, the O.D.E. for the \textit{total recovered people} $R_t$ (i.e. the total individuals having survived the disease) is trivially obtained by considering that the rate of $R_t$ is approximatively proportional to the number of the infected people $n_t$ at time $t$ i.e.\footnote{Notice that Eq.~(\ref{dp1b}) is the dynamic equation for the total recovered people adopted in the \textit{Susceptible-Infectious-Recovered-Deceased-Model} (SIRD-model) \cite{kermack}. The comparison between our model with the SIRD-model will be found soon in Ref.~\cite{sonninoM}.}.
\begin{equation}\label{dp1b}
\frac{dR_t}{dt}=\chi n_t
\end{equation}
\noindent However, it is useful to clarify the following. In Eqs~(\ref{dp1a}), $h_t$ stands for the \textit{total number of the recovered people previously hospitalised} whereas the variable $R_t$ in Eq.~(\ref{dp1b}) is the \textit{total number of the recovered people} (i.e. the number of the recovered people previously hospitalised, plus the number of the asymptomatic people, plus the infected people who have been recovered without being previously hospitalised). The natural question is: \textit{"how can we count $R_t$ and compare this variable with the real data ?"}. The current statistics, produced by the Ministries of Health of various Countries, concern the people released from the hospitals. Apart from Luxembourg (where the entire population has been subject to the COVID-19-test), no other Countries are in a condition to provide statistics regarding the total people recovered by COVID-19. Hence, it is our opinion that the equation for $R_t$, is not useful since it is practically impossible to compare $R_t$ with the experimental data. 
\subsection{Equation for the Deceased People}
\noindent The rate of deceased people per unit time is modelled by the following dimensionless equation
\begin{equation}\label{dp2}
\frac{d}{d{\hat t}}d_t=\alpha_d n_{(t-t_d)}-\beta_dn_{(t-t_d)}^2
\end{equation}
\noindent The meaning of Eq.~(\ref{dp2}) is the following. Manifestly, the rate of deaths is proportional to the number of active people. However, individuals infected by SARS-CoV-2 do not die instantly since the rate of deaths at time $t$ is proportional to the people who were infected at an earlier time $t-t_d$ ($ t_d >0 $) with $t_d$ denoting the {\it time-delay}. We indicate with $\alpha_d$ the, time-independent, constant proportional to the increase of the deaths $d_t$. The second term, $-\beta_dn_{(t-t_d)}^2$, models the presence of the lockdown measures, having the effect of saturating the rate of infected people and, consequently, of deaths. Indeed, in the absence of lockdown measures, we may approximatively write
\begin{equation}\label{dp2a}
\frac{d}{d{\hat t}}d_t=\alpha_r n_{(t-t_d)}
\end{equation}
\noindent with $\alpha_r$ denoting a positive constant. The purpose of the lockdown measures is to decrease the number of infected people, and therefore deaths. We may assume that the effect of the lockdown measures is proportional to $n_{(t-t_d)}$ such as to dampen the linear growth of the mortality rate. In other terms, we get 
\begin{equation}\label{dp2b}
\alpha_r \rightarrow \alpha_r -\beta_rn_{(t-t_d)}
\end{equation}
\noindent which combined with Eq.~(\ref{dp2a}) gives Eq.~(\ref{dp2}).

\subsection{Equation for the Positive People}
\noindent Of course, during the descent phase, the number of active people $n_t$ satisfies a simple law of conservation:  \textit{If we are in the situation where there are no longer new cases of people tested positive for COVID-19 and if we assume that the active people cannot leave their country of origin (or else, if they do, they will be rejected by the host Country), then the number of infected people cannot but decrease either because some people are deceased or because others have been recovered}. In mathematical terms
\begin{equation}\label{dp4}
n_t=n_{0}-(h_t-h_0)-(d_t-d_0)=N_{Max}-h_t-d_t
\end{equation}
\noindent with $h_0$, $d_0$ and $n_0$ denoting the values of $h_t$, $d_t$ and $n_t$ evaluated at the time $t=t_{Max}$ (see Eq.~(\ref{M10}) i.e., the time that maximises the number of the total cases), respectively. It should be noted that the conservation law~(\ref{dp4}) applies only when there are no longer new cases of people tested positive to COVID-19\footnote{So, Eq.~(\ref {dp4}) does not apply necessarily as soon as the number $n_t$ (the number of people tested positive for COVID-19) starts to decrease. Indeed, it may happen that $n_t$ decreases because, for example, the number of new cases of people tested positive is less than the number of the people who have recovered in the meantime. Conservation law~(\ref {dp4}) applies only from the moment where the number of new cases of people tested positive is strictly equal to zero.}. Here, by the {\it descending phase} we mean the phase where Eq.~(\ref {dp4}) applies. To sum up, the equations describing the descending phase are
\begin{align}\label{dp5}
&\frac{d}{d{\hat t}}r_t=\alpha_r r_t\Bigl(1-\frac{r_t}{K_r}\Bigr)\qquad\qquad\qquad\quad\quad{\rm with}\ \ r_{t=t_{Max}}=r_0\\
&\frac{d}{d{\hat t}}d_t=\alpha_d n_{(t-t_d)} -\beta_dn_{(t-t_d)}^2 \qquad\qquad\quad\ \!{\rm with}\ \ d_{t=t_{Max}}=d_0\nonumber\\
&n_t=N_{Max}-h_t-d_t\qquad\qquad\qquad\qquad\ \ {\rm with}\ \ n_{\infty}=0\nonumber\\
&h_t=\sum_{n=1}^{n=t/\Delta t} r_n\qquad\qquad\qquad\qquad\qquad\quad\ \ \!{\rm where}\ \  \Delta t\simeq 1\ {\rm day}\nonumber
\end{align}
\noindent with $t_{Max}$ given by Eq.~(\ref{M10}). Notice that the first two equations of system~(\ref{dp5}) are valid also during the {\it ascending-phase}. Of course, in this case, the initial conditions are $r_{t=0}=0$, $d_{t=0}=0$ and $n_{t=0}=0$. Hence, during the ascending phase the evolution equations are
\begin{align}\label{dp6}
&\frac{d}{d{\hat t}}r_t=\alpha_r r_t\Bigl(1-\frac{r_t}{K_r}\Bigr)\qquad\qquad\qquad\quad\quad{\rm with}\ \ r_{t=0}=0\\
&\frac{d}{d{\hat t}}d_t=\alpha_d n_{(t-t_d)} -\beta_dn_{(t-t_d)}^2\qquad\qquad\quad {\rm with}\ \ d_{t=0}=0\nonumber\\
&n_t=N_t-h_t-d_t\qquad\qquad\qquad\qquad\qquad {\rm with}\ \ n_{t=0}=0\nonumber\\
&\frac{d}{d{\widehat t}}N_t=\alpha N_t\Bigl(1-\frac{N_t}{K_N}\Bigr)-\Bigl(\frac{\alpha {\widehat t}^2- 1}{{\widehat t}}\Bigr)N_t\nonumber\\
&h_t=\sum_{n=1}^{n=t/\Delta t} r_n\qquad\qquad\qquad\qquad\qquad\quad\ \ \!{\rm where}\ \  \Delta t\simeq 1\ {\rm day}\nonumber
\end{align}
According to our expectations, by solving numerically system~(\ref{dp6}), with good approximation, we get
\begin{equation}\label{dp5bis}
d_t\propto n_{(t-t_d)}
\end{equation}
\noindent
\noindent Next, we find the numerical solution of systems~(\ref{dp5})-(\ref{dp6}) for Italy and Belgium. A similar analysis for Luxembourg is reported in Appendix.
\subsection{Theoretical Predictions for the Descending Phase}
In this subsection, we report the numerical solutions of Eqs~(\ref{dp5})-(\ref{dp6}) for Italy and Belgium. The  solution for Luxembourg can be found in the Appendix. Fig.~(\ref{Fig.12}) and (\ref{Fig.13}) concern the Italian situation. They show the numerical solution of Eqs~(\ref{dp5})-(\ref{dp6}) for the number of recovered people and deaths, respectively. These theoretical predictions are plotted against the experimental data reported in the Table~(\ref{A}). According to the theoretical predictions, for Italy we get $t_{ITd}=12$ days. Fig.~(\ref{Fig.14}), illustrates the descendant-phase for Italy.
\begin{figure*}[t]
  \hfill
    \begin{minipage}[t]{.45\textwidth}
    \begin{center}
      \includegraphics[width=5cm,height=5cm]{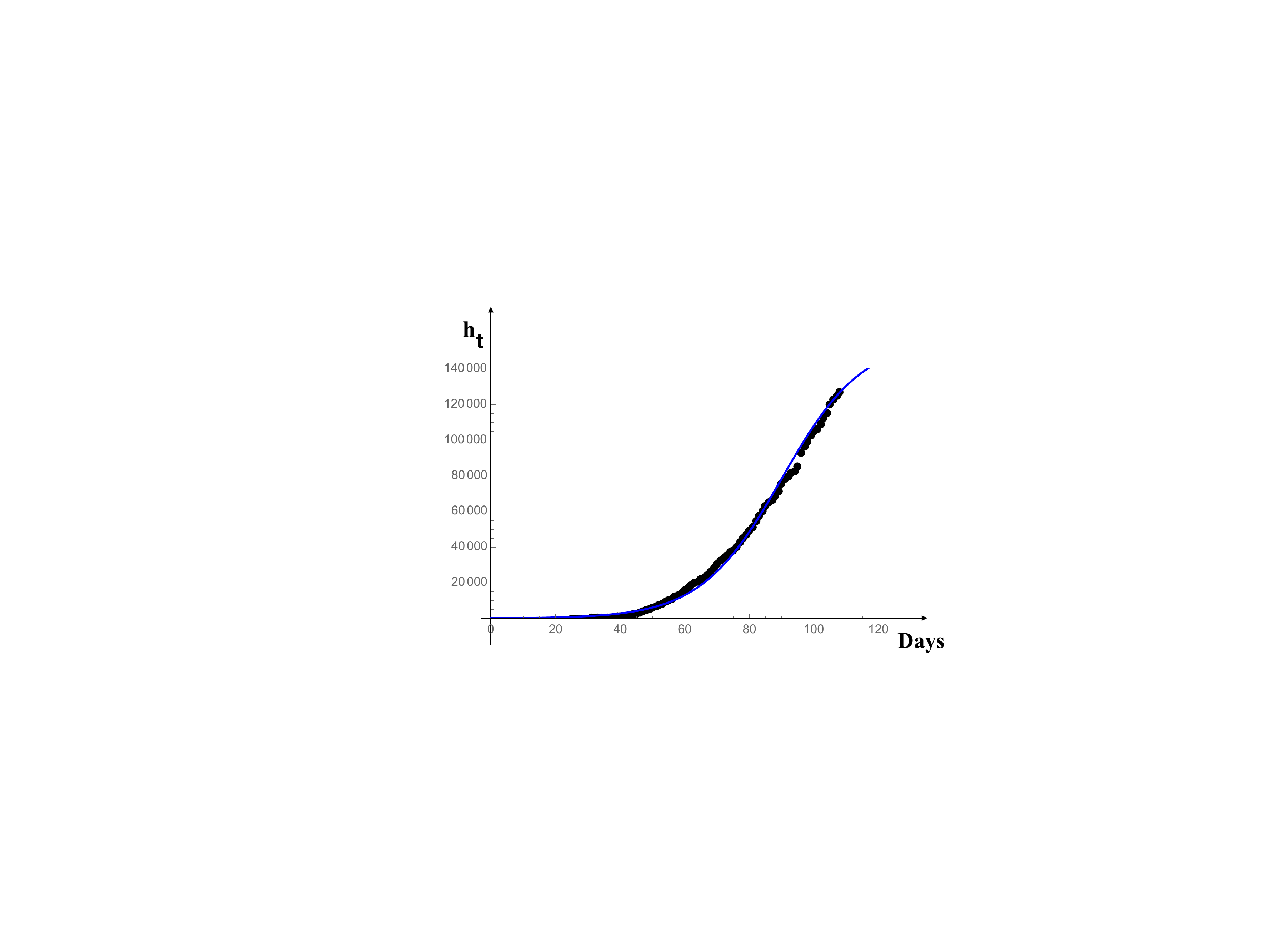}
      \caption{\textit{Italy situation}. Theoretical predictions (blue line) against the experimental data (black circles) for the recovered people.}
      \label{Fig.12}
    \end{center}
  \end{minipage}
  \hfill
  \begin{minipage}[t]{.45\textwidth}
    \begin{center}
      \includegraphics[width=5cm,height=5cm]{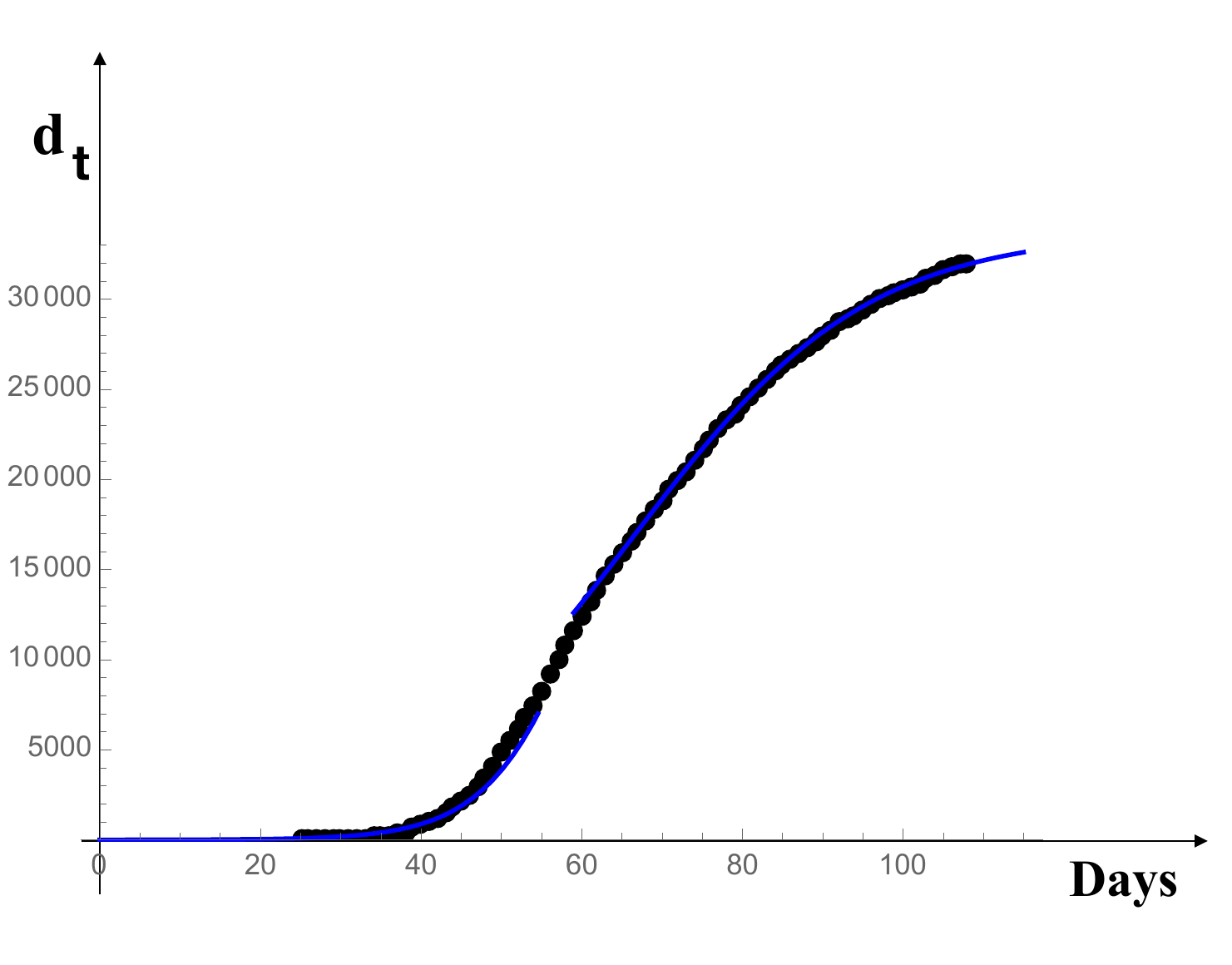}
      \caption{\textit{Italy situation}. Theoretical predictions (blue line) against the experimental data (black circles) for the deceased people.}
      \label{Fig.13}
    \end{center}
  \end{minipage}
  \hfill
\end{figure*}
\begin{figure}[th!]
\hskip 2truecm
\includegraphics[width=8cm,height=8cm]{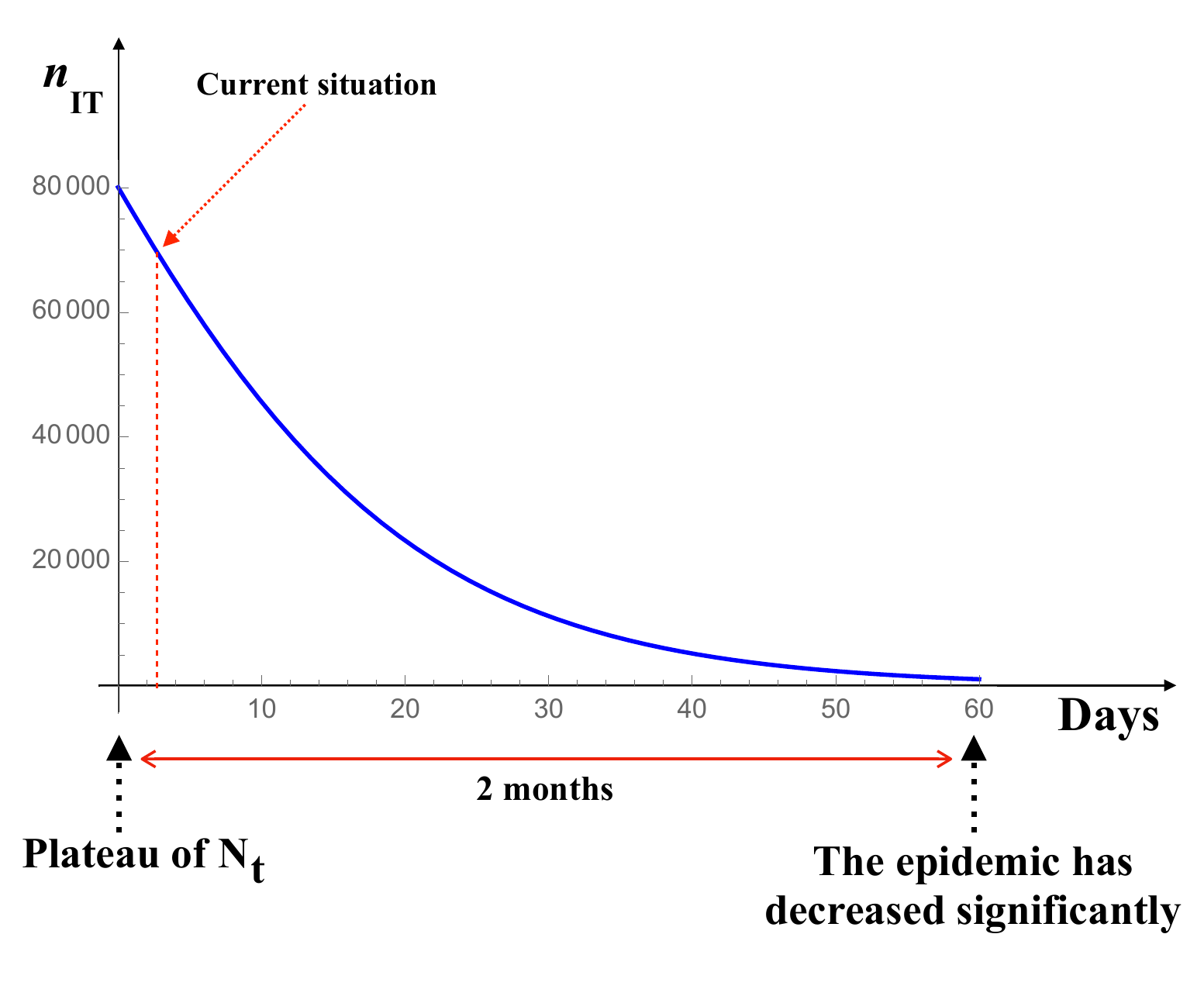}
\caption{\textit{The descending phase for Italy}. According to the theoretical predictions, after two months the lockdown measures may heavily be lightened and we can return to normal work. The estimated {\textit time-delay} is $t_{ITd}=12$ days - see Eq.~(\ref{dp5})}
\label{Fig.14}
\end{figure}

\noindent Figs~(\ref{Fig.15}) and (\ref{Fig.16}) refer to the Belgian situation. The figures illustrate the numerical solutions of Eqs~(\ref{dp5})-(\ref{dp6}) for the number of recovered people and deaths, respectively. The theoretical predictions are plotted against the experimental data reported in the Table~(\ref{B}). According to the theoretical predictions, for Belgium we get $t_{BEd}=8.8$ days. Fig.~(\ref{Fig.17}) shows the descendant-phase for Belgium.
\begin{figure*}[t]
  \hfill
    \begin{minipage}[t]{.45\textwidth}
    \begin{center}
      \includegraphics[width=5cm,height=5cm]{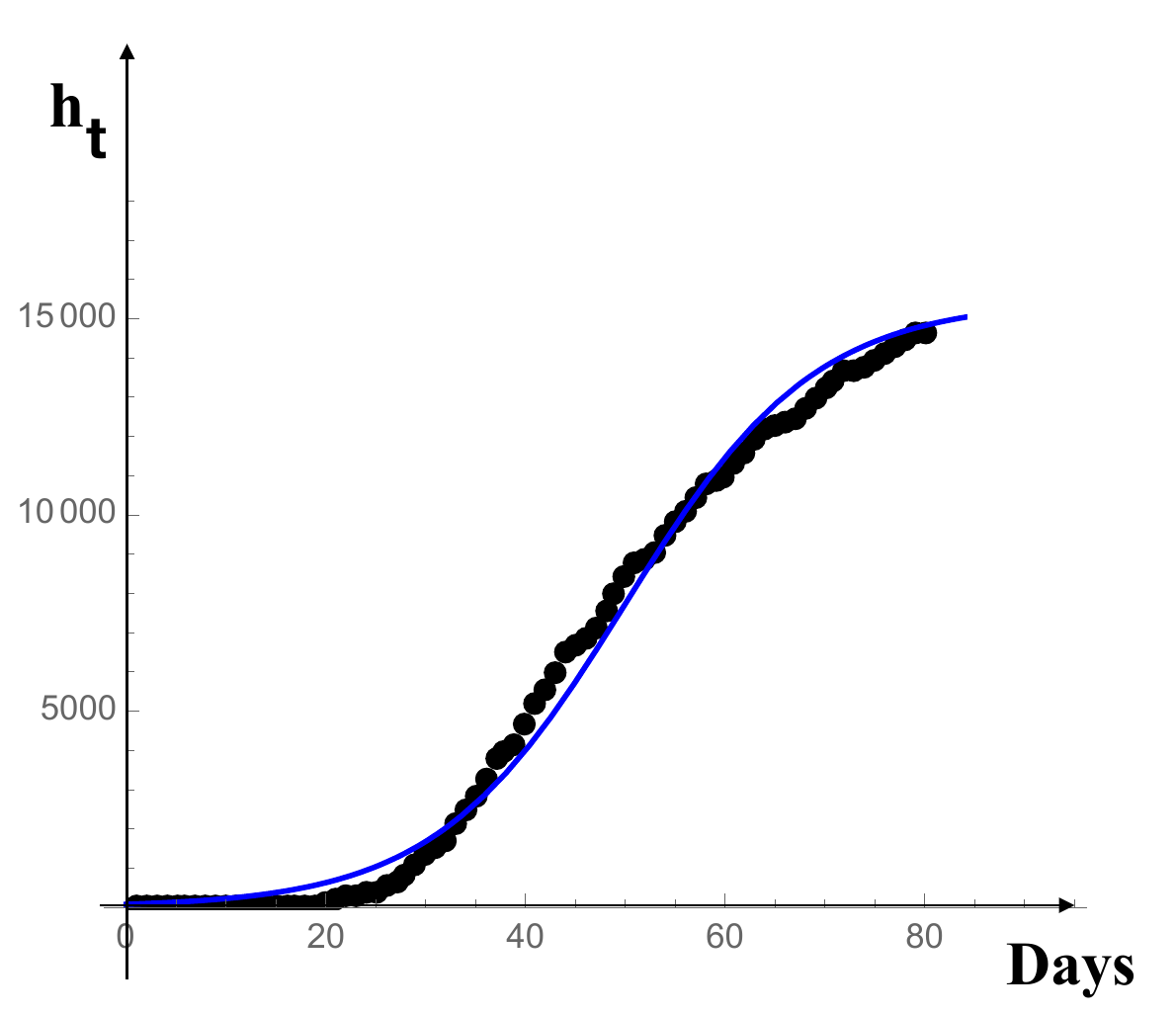}
      \caption{\textit{Belgian situation}. Theoretical predictions (blue line) against the experimental data (black circles) for the recovered people.}
      \label{Fig.15}
    \end{center}
  \end{minipage}
  \hfill
  \begin{minipage}[t]{.45\textwidth}
    \begin{center}
      \includegraphics[width=5cm,height=5cm]{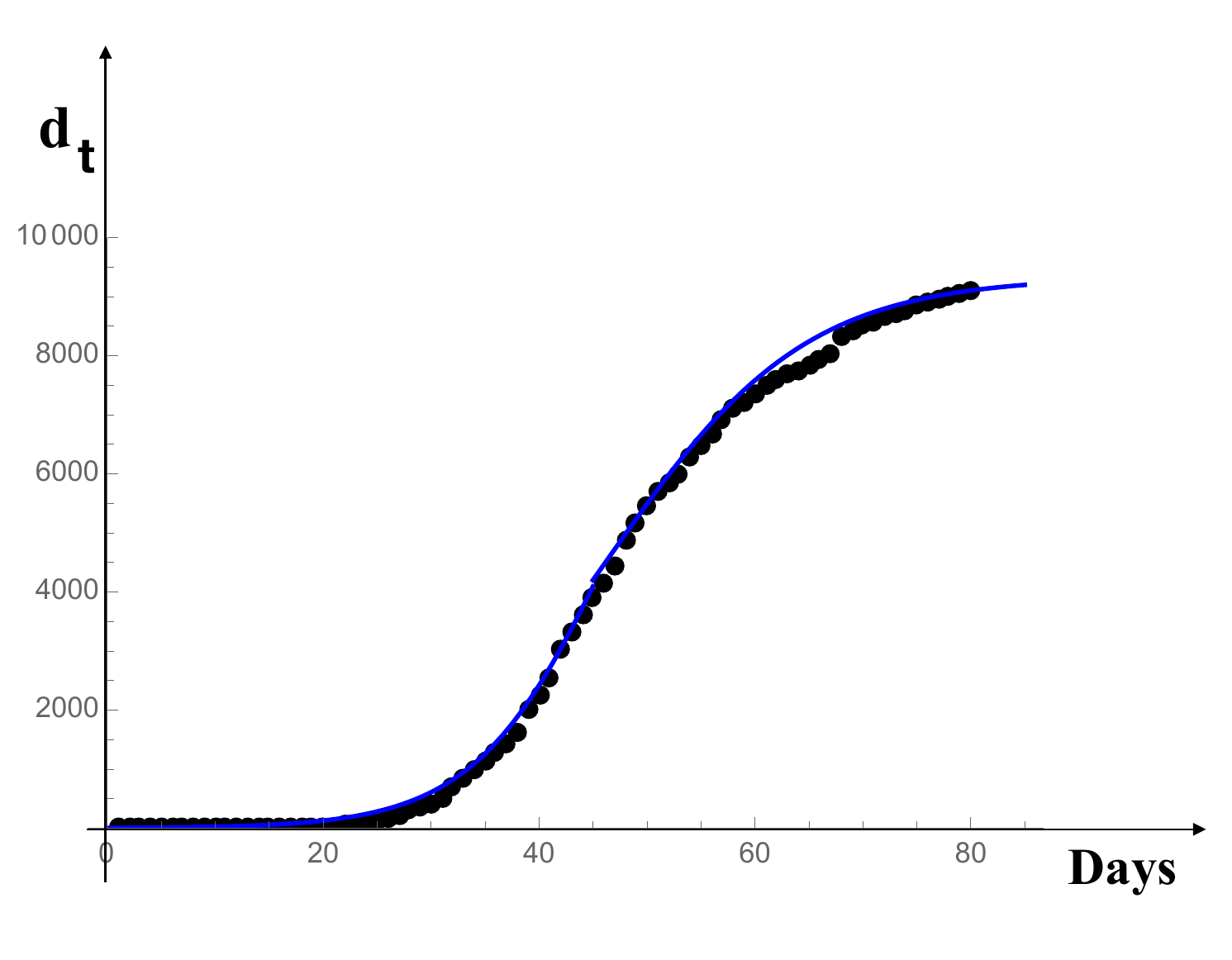}
      \caption{\textit{Belgian situation}. Theoretical predictions (blue line) against the experimental data (black circles) for the deceased people.}
      \label{Fig.16}
    \end{center}
  \end{minipage}
  \hfill
\end{figure*}
\begin{figure}[th!]
\hskip 2.5truecm
\includegraphics[width=7cm, height=7cm]{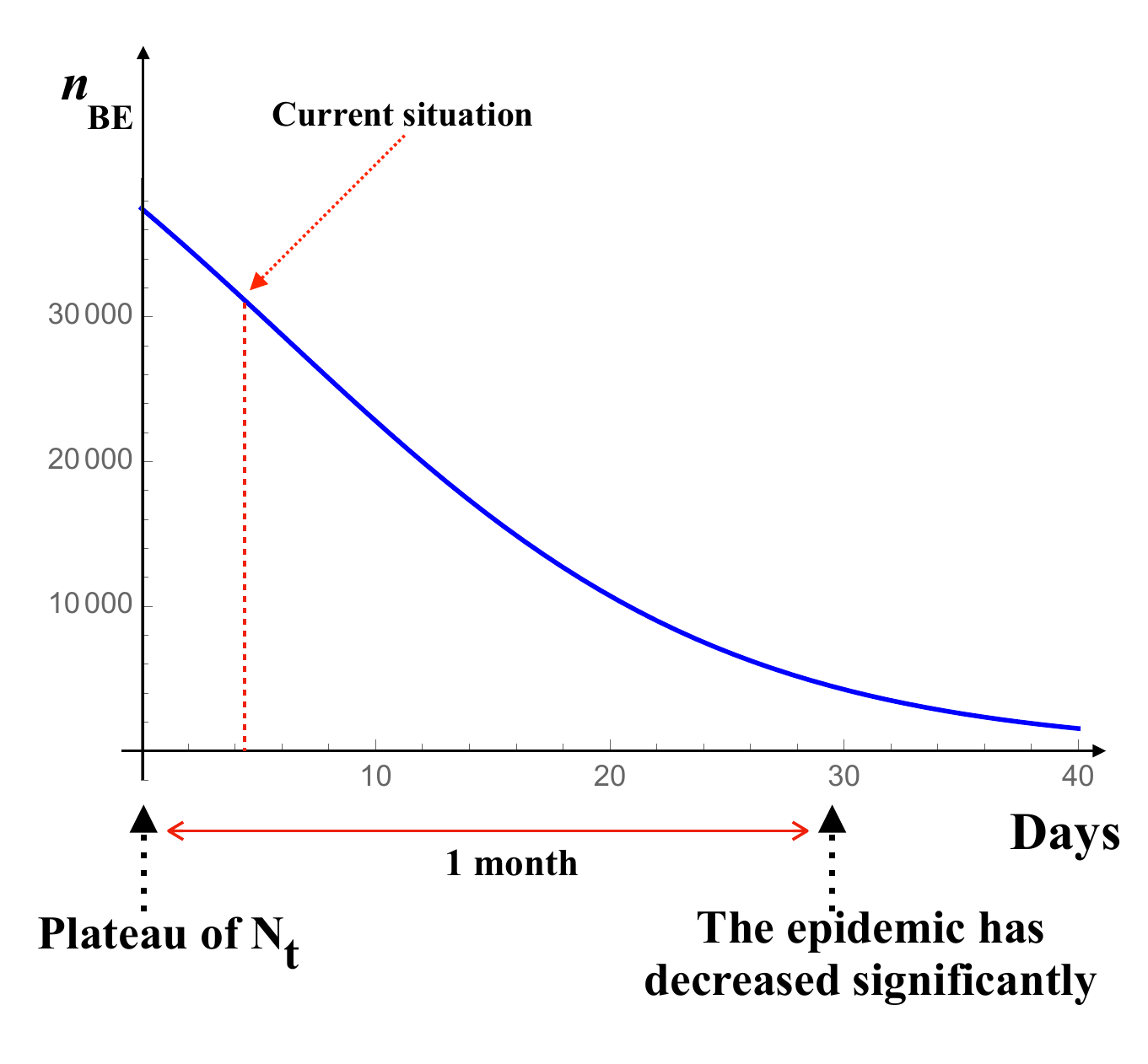}
\caption{\textit{The descending phase for Belgium}. According to the theoretical predictions, after one month the lockdown measures may heavily be lightened and we can return to normal work. The estimated {\textit time-delay} is $t_{BEd}=8.8$ days - see Eq.~(\ref{dp5}).}
\label{Fig.17}
\end{figure}

\section{Conclusions}\label{conclusions}
In this work we studied the spread of SARS-CoV-2 until when the strict lockdown measures have been adopted (i.e. until 16th May 200). The dynamics of COVID-19 when the population is under less restrictive lockdown measures will be subject of future studies. Through fitting techniques previously performed, and validated by the $\chi^2$-tests, we determined a general differential equation, having two-free parameters, which we propose for 13 respiratory infectious diseases caused by viruses, including SARS-CoV-2. The solution of Eq.~(\ref{MN9}) provides the number of the total case in time. Successively, we compared the theoretical predictions, provided by the solution of Eq.~(\ref{MN9}) and by the logistic model (see Eq.~(\ref{M1})), with the real data for Italy and Belgium (for Luxembourg see Appendix). We saw that the solution of Eq.~(\ref{MN9}) is in good agreement with the experimental data since the beginning of the appearance of the COVID-19; this is not the case for the logistic model which applies only to the few last days.
We found the days where the maximum number infected people by COVID-19 will be reached in Italy and Belgium by parametrising the solution of Eq.~(\ref{MN9}) with experimental data: we get, $t_{MaxIT}\simeq 21\ \textrm{ April}\ 2020$ and $t_{MaxBE}\simeq 2\ \textrm{ May}\ 2020$ for Italy and Belgium, respectively.

\noindent We also noted, empirically, that the infection process caused by SARS-CoV-2 may be divided into three qualitatively different periods; i.e., the \textit{exponential period}, the \textit{transient period} and the \textit{bell-shaped period} (or the \textit{post-transient period}). The solution of Eq.~(\ref {MN9}) allows defining more precisely these three periods. Indeed, we may classify the above periods as follows
\begin{align}\label{conclusions1}
&\textrm{ The\ exponential\ period}\quad\quad\ \!\textrm{ for}\quad 0\leq {\widehat t}\leq {\widehat t}_{LM}\\
&\textrm{ The\ transient\ period}\quad\quad\quad\ \textrm{ for}\quad\! {\widehat t}_{LM}< {\widehat t} \leq {\widehat t}_{flex}\nonumber\\
&\textrm{ The\ bell-shaped\ period}\quad\quad\  \textrm{ for}\ \ {\widehat t}> {\widehat t}_{flex}\nonumber
\end{align}
\noindent with ${\widehat t}_{LM}$ indicating the {\it dimensionless time when the lockdown measures are applied} and ${\widehat t}_{flex}$ the \textit{dimensionless inflection point of the solution of} Eq.~(\ref{MN9}), respectively. It is easily checked that, for large values of $K_N$, the value of ${\hat t}_{flex}$ satisfies, approximatively, the equation
\begin{equation}\label{conclusions2}
\alpha {\widehat t}_{flex}^3-2\alpha{\widehat t}_{flex}^2+(\alpha-3){\widehat t}_{flex}+2\simeq 0\quad{\rm with}\ \ {\widehat t}_{flex}\equiv\frac{{t}_{flex}}{t_0}
\end{equation}
\noindent Hence, according to Eq.~(\ref{conclusions2}), the \textit{transient period} ended on 31 March 2020 for Italy and on 7 April 2020 for Belgium, respectively. The second part of the work is devoted to modelling the descending phase, i.e. the decrease of the number of people tested positive for COVID-19. Also in this case, we proposed a new set of dynamic differential equations that we solved numerically. The solution of Eqs~(\ref{dp5}) (and Eq.~(\ref{dp6})) provided valuable information such as the duration of the COVID-19 epidemic in a given Country and therefore when it will be possible to return to normal life.

\section{Acknowledgments}\label{ack}
I am very grateful to Alberto Sonnino from Facebook Calibra and University College London for comments on late manuscript, and to Ing. Alessandro Leone from the Italian Embassy in Belgium for his suggestions and encouragement.

\section{Appendix: Comparison between the Theoretical Predictions of Eq.~(\ref{MN9}) and Experimental Data for Luxembourg}\label{app}
Luxembourg, due to the particularly severe lockdown measures adopted by the government, may be considered, with good approximation, as a closed system (628108 inhabitants, most of them concentrated in only one town). Indeed, right from the start, the city of Luxembourg was literally closed and citizens were unable to enter and leave the city freely (people who had to enter in the city for working reasons were obliged to undergo each time the test that, of course, had to result negative). 

\noindent Italy, on the other hand may be considered, with a good approximation, as an open system (60317116 inhabitants dislocated in all the Country). In Italy, especially during the initial phase, the citizens of northern Italy moved freely to the south of Italy, by train, by plans or by car. Only in a second time the Italian government decided to introduce much more restrictive measures concerning the movement of citizens from one region to another. 

\noindent For the reason mentioned above, it is our opinion that it is very interesting to analyse these two Countries, Luxembourg and Italy, which are so different with each other. In this Appendix we report the comparison between the theoretical predictions of the COVID-19 model~(\ref{MN9}) and the real data for Luxembourg update to 15 May 2020 (see Fig.~(\ref{Fig.18})). In the columns of table~(\ref{C}) we can find the number of active people (currently infected by SARS-CoV-2), the number of recovered people, and the number of deceased people, respectively. The experimental data have been found in the databases \cite{Luxdata1} and \cite{Luxdata2}. Luxembourg reached its peak on 12 April 2020. 
\input{Data_LUX.tex}
\input{Data_LUX_II.tex}
\begin{figure}
\hskip 2.5truecm
\includegraphics[width=7cm, height=7cm]{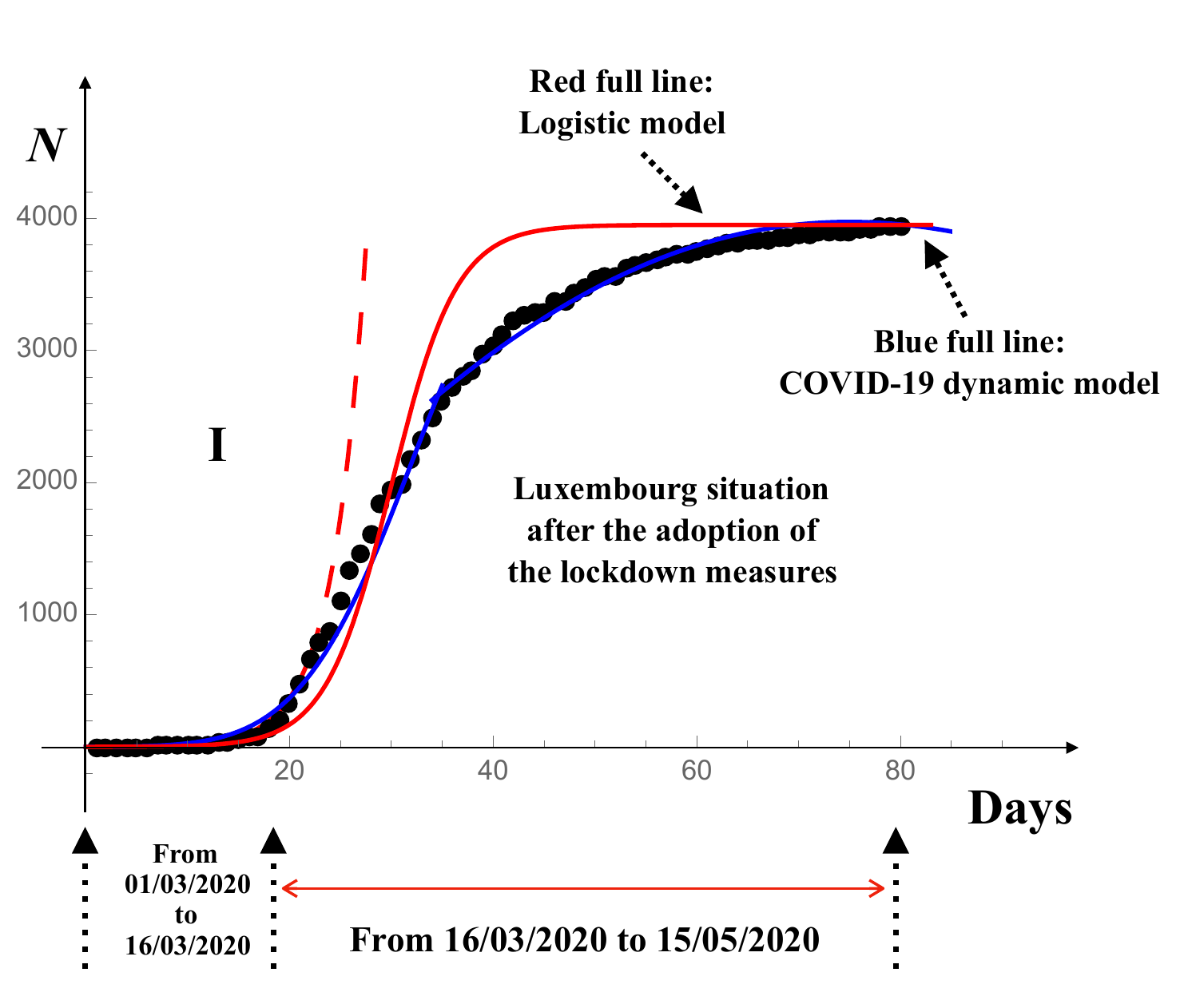}
\caption{\textit{Situation in Luxembourg on 15 May 2020}. The black dots correspond to real data. The red dotted line corresponds to the situation in Luxembourg before the adoption of the lockdown measures. The blue and the red solid lines correspond to the theoretical predictions for Luxembourg according to the solution of Eq.~(\ref{MN9}) and the logistic model, respectively. Solution of Eq.~(\ref{MN9}) fits well all the experimental data from the initial days (i.e., from the 29th of February 2020), while the logistic model applies only to the first data. The values of the parameters of Eq.~(\ref{MN9}) and the logistic function~(\ref{M1c}) are: $\tau_{LUX}\simeq 3.2$ days ($\mu_{LUX}=2.2$ days), $K_N^{LUX}\simeq 6880$, and $t_{0LUX}\simeq 40$ days for Eq.~(\ref{MN9}), and $\tau_{LUX}\simeq 3.2$ days ($\mu_{LUX}=2.2$ days), $K_{LUX}=3950$, $t_{0LLUX}=30$ days for the Logistic function, respectively. The zone $I$ corresponds to the period before the adoption of the lockdown measures.}
\label{Fig.18}
\end{figure}

\subsection{The Descending Phase for Luxembourg}
Figs~(\ref{Fig.19}) and (\ref{Fig.20}) refer to the Luxembourg situation. The figures illustrate the numerical solutions of Eqs~(\ref{dp5})-(\ref{dp6}) for the number of recovered people and deaths, respectively. The theoretical predictions are plotted against the experimental data, which can be found in the Table~(\ref{C}). Fig.~(\ref{Fig.21}) shows the descending phase for Luxembourg.

\begin{figure*}[t]
  \hfill
    \begin{minipage}[t]{.45\textwidth}
    \begin{center}
      \includegraphics[width=5cm,height=5cm]{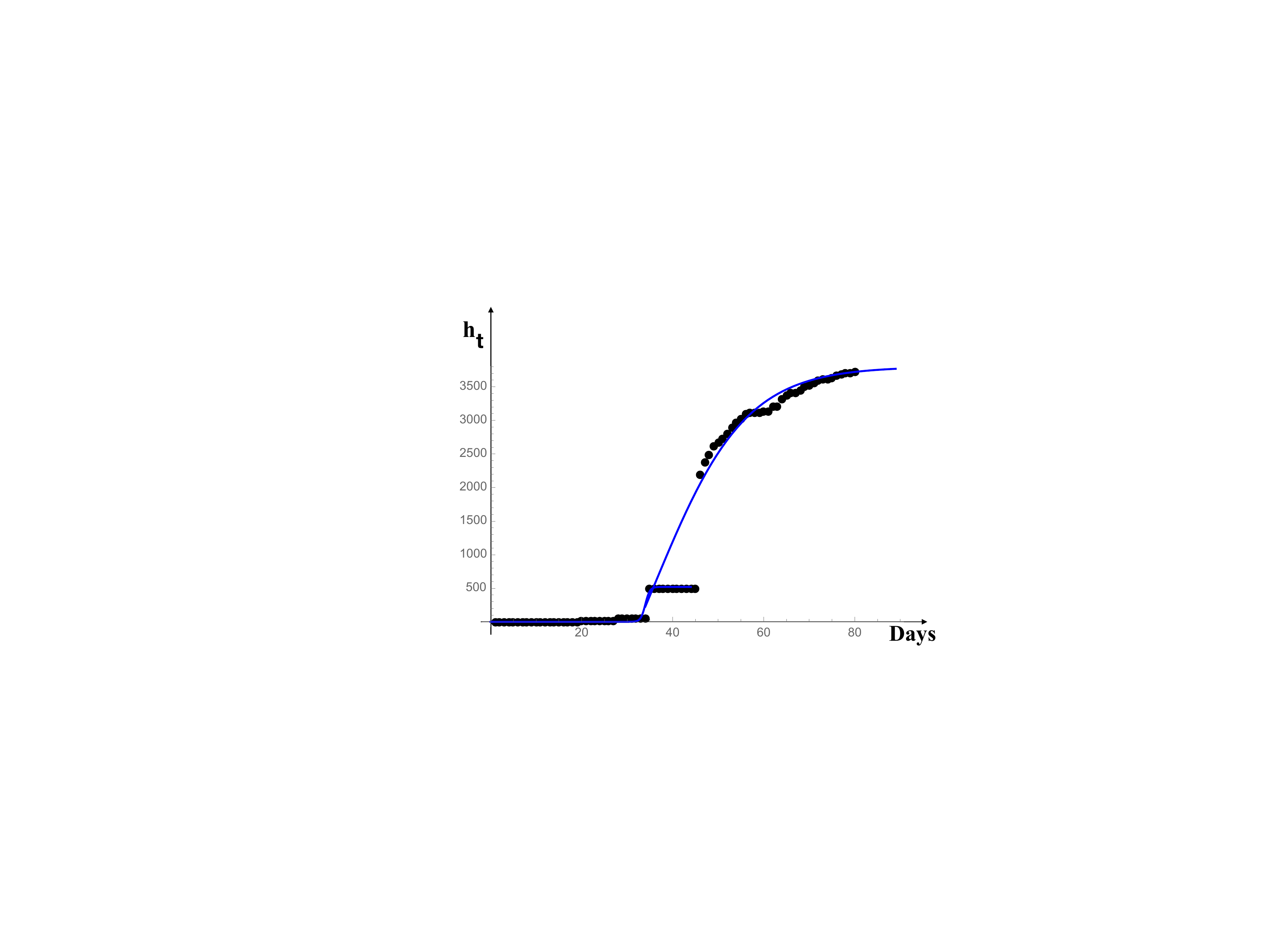}
      \caption{\textit{Luxembourg situation}. Theoretical predictions (blue line) against the experimental data (black circles) for the recovered people.}
      \label{Fig.19}
    \end{center}
  \end{minipage}
  \hfill
  \begin{minipage}[t]{.45\textwidth}
    \begin{center}
      \includegraphics[width=5cm,height=5cm]{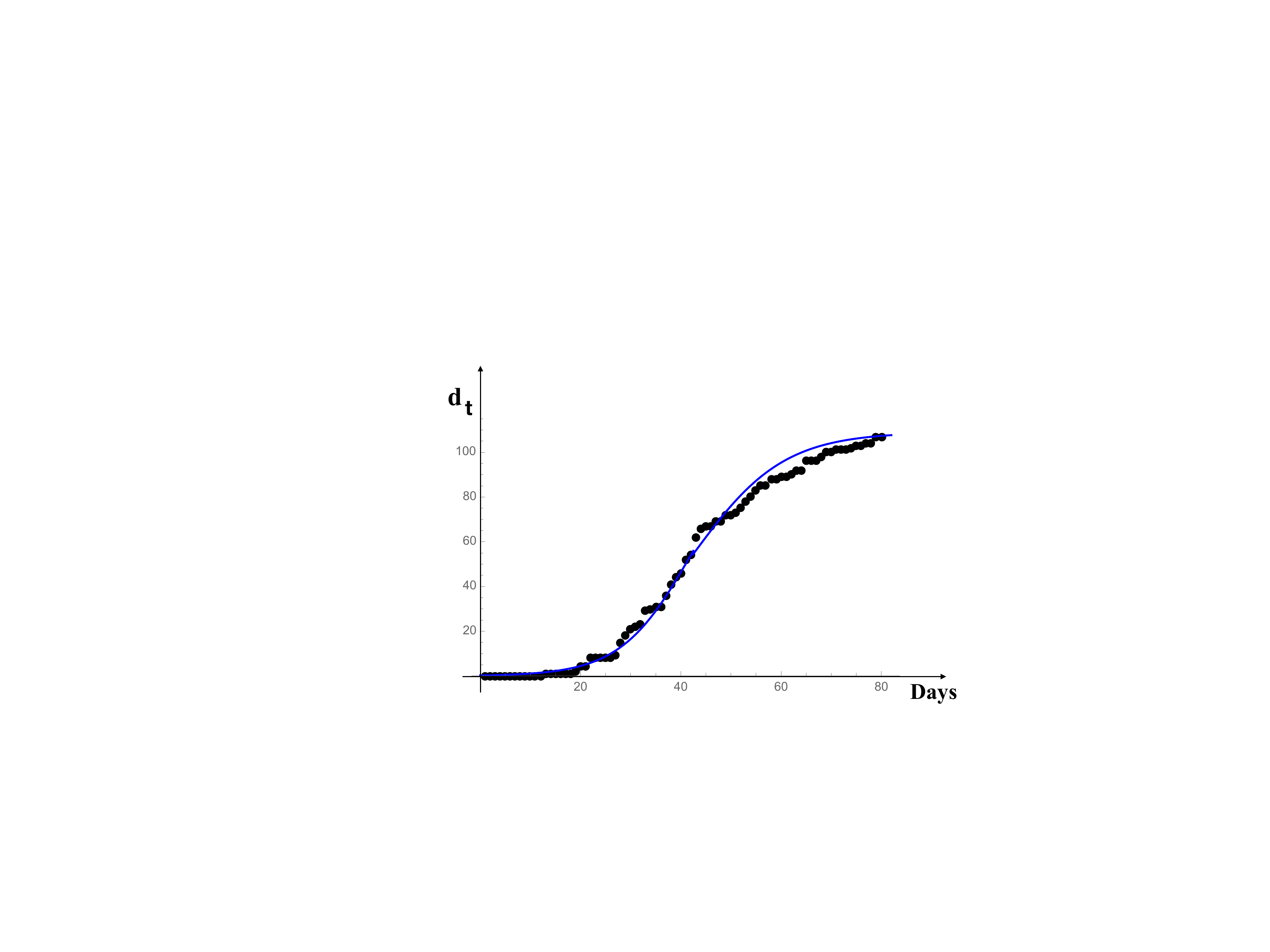}
      \caption{\textit{Luxembourg situation}. Theoretical predictions (blue line) against the experimental data (black circles) for the deceased people.}
      \label{Fig.20}
    \end{center}
  \end{minipage}
  \hfill
\end{figure*}
\begin{figure}[th!]
\hskip 2.5truecm
\includegraphics[width=7cm, height=7cm]{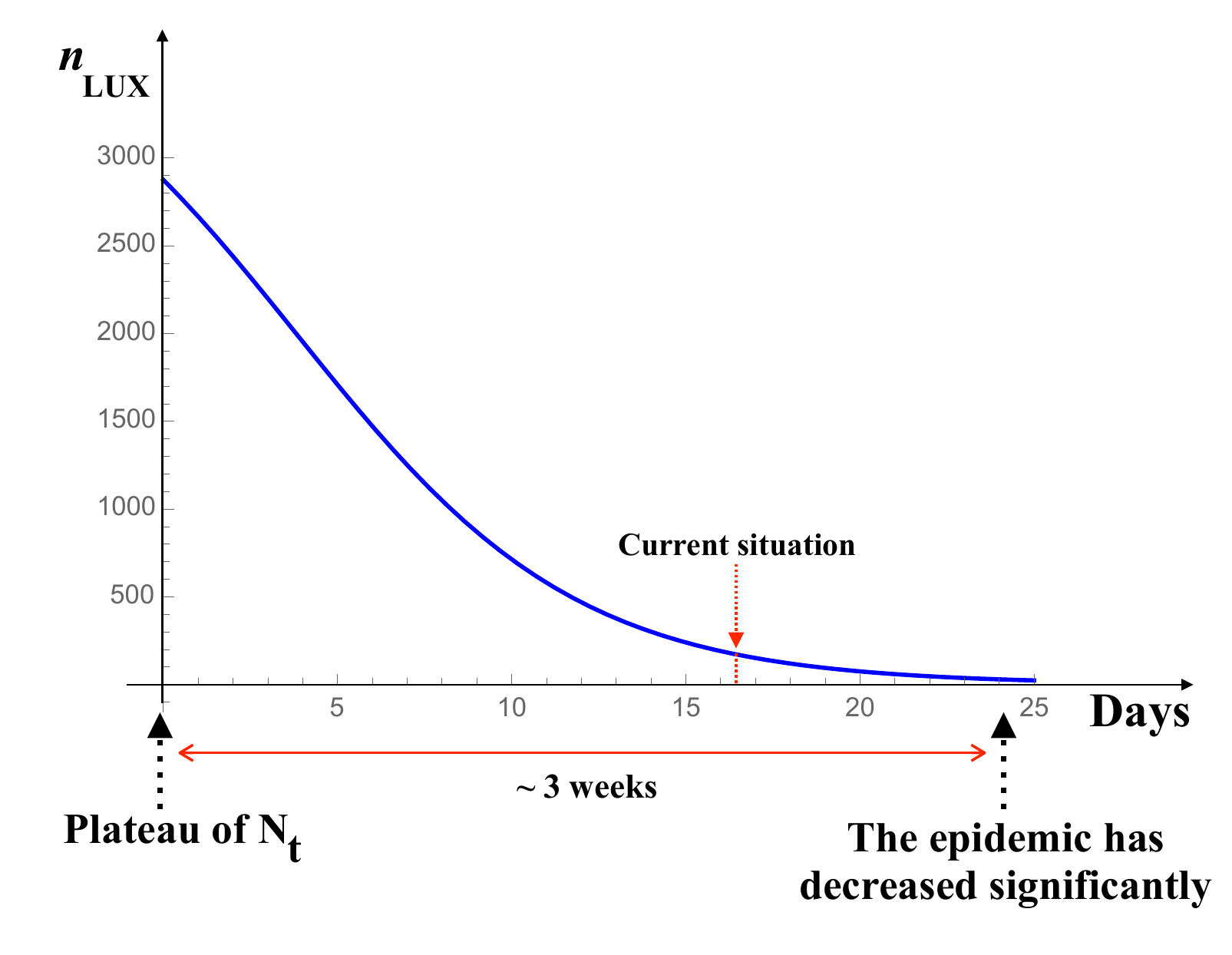}
\caption{\textit{The descending phase for Luxembourg}. According to the theoretical predictions, after one month the lockdown measures may heavily be lightened and we can return to normal work. The estimated {\textit time-delay} is $t_{LUXd}=15$ days - see Eq.~(\ref{dp5}).}
\label{Fig.21}
\end{figure}

\end{document}

%% file: Data_IT.tex
\begin{table}[t]
\centering
\footnotesize
\begin{tabular}{lrrrr}
  \toprule
  Date & Active & Recovered & Deceased & Total cases\\
  \midrule
  25-Feb & 322 & 1 & 10 & 333 \\
  26-Feb & 400 & 3 & 12 & 415 \\
  27-Feb & 650 & 45 & 18 & 713 \\
  28-Feb & 888 & 46 & 21 & 955 \\
  29-Feb & 1049 & 50 & 29 & 1128 \\
  01-Mar & 1577 & 83 & 34 & 1694 \\
  02-Mar & 1835 & 149 & 52 & 2036 \\
  03-Mar & 2263 & 160 & 79 & 2502 \\
  04-Mar & 2706 & 276 & 107 & 3089 \\
  05-Mar & 3296 & 414 & 148 & 3858 \\
  06-Mar & 3916 & 523 & 197 & 4636 \\
  07-Mar & 5061 & 589 & 233 & 5883 \\
  08-Mar & 7375 & 622 & 366 & 8363 \\
  09-Mar & 9172 & 724 & 463 & 10359 \\
  10-Mar & 10149 & 1004 & 631 & 11784 \\
  11-Mar & 10590 & 1045 & 827 & 12462 \\
  12-Mar & 12839 & 1258 & 1016 & 15113 \\
  13-Mar & 14955 & 1439 & 1266 & 17660 \\
  14-Mar & 17750 & 1966 & 1441 & 21157 \\
  15-Mar & 20603 & 2335 & 1809 & 24747 \\
  16-Mar & 23073 & 2749 & 2158 & 27980 \\
  17-Mar & 26062 & 2941 & 2503 & 31506 \\
  18-Mar & 28710 & 4025 & 2978 & 35713 \\
  19-Mar & 33190 & 4440 & 3405 & 41035 \\
  20-Mar & 37860 & 5129 & 4032 & 47021 \\
  21-Mar & 42681 & 6072 & 4825 & 53578 \\
  22-Mar & 46638 & 7024 & 5475 & 59137 \\
  23-Mar & 50418 & 7432 & 6077 & 63927 \\
  24-Mar & 54030 & 8326 & 6820 & 69176 \\
  25-Mar & 57511 & 9362 & 7503 & 74376 \\
  26-Mar & 62013 & 10361 & 8165 & 80539 \\
  27-Mar & 66414 & 10950 & 9134 & 86498 \\
  28-Mar & 70065 & 12384 & 10023 & 92472 \\
  29-Mar & 73880 & 13030 & 10779 & 97689 \\
30-Mar & 75528 & 14620 & 11591 & 101739 \\
  31-Mar & 77635 & 15729 & 12428 & 105792 \\
  01-Apr & 80572 & 16847 & 13155 & 110574 \\
  02-Apr & 83049 & 18278 & 13915 & 115242 \\
  03-Apr & 85388 & 19758 & 14681 & 119827 \\
  04-Apr & 88274 & 20996 & 15362 & 124632 \\
  05-Apr & 91246 & 21815 & 15887 & 128948 \\
  06-Apr & 93187 & 22837 & 16523 & 132547 \\
  07-Apr & 94067 & 24392 & 17127 & 135586 \\
  08-Apr & 95362 & 26491 & 17669 & 139422 \\
   09-Apr & 96877 & 28470 & 18279 & 143626 \\
   10-Apr & 98273 & 30455 & 18849 & 147577 \\
   11-Apr & 102253 & 34211 & 19899 & 152271 \\
   12-Apr & 100269 & 32534 & 19468 & 156363\\
   13-Apr & 103616 & 35435 & 20465 & 159516\\
   14-Apr & 104291 & 37130 & 21067 & 162488\\
   15-Apr & 105418 & 38092 & 21645 & 165155\\
  \bottomrule
\end{tabular}
\caption{\footnotesize\textit{Situation in Italy on 15 May 2020}. Columns report the number of active people (currently infected by SARS-CoV-2), the number of recovered people, and the number of deceased people.}
\label{A}
\end{table}

%% file: Data_IT_II.tex
\begin{table}[t]
\centering
\footnotesize
\begin{tabular}{lrrrr}
  \toprule
  Date & Active & Recovered & Deceased & Total cases\\
  \midrule
  16-Apr & 106607 & 40164 & 22170 & 168941 \\
  17-Apr & 106962 & 42727 & 22745 & 172434 \\
 18-Apr & 107771 & 44927 & 23227 & 175925 \\
  19-Apr & 108257 & 47055 & 23660 & 178972 \\
  20-Apr &  108237&  48877 & 24114 & 181228 \\
  21-Apr & 107709 & 51600 & 24648 & 183957 \\
  22-Apr & 107699 & 54543 & 25085 & 187327 \\
  23-Apr  & 106848 & 57576 & 25549 & 189973 \\
  24-Apr  & 106527 & 60498 & 25969 & 192994 \\
  25-Apr  & 105847 & 63120 & 26348 & 195351 \\
  26-Apr  & 106103 & 64928 & 26644 & 197675 \\
  27-Apr  & 105813 & 66624 & 26977 & 199414 \\
  28-Apr  & 105205 & 68941 & 27359 & 201505 \\
  29-Apr  & 104657 & 71252 & 27682 & 203591 \\
  30-Apr  & 101551 & 75945 & 27967 & 205463 \\
  01-May & 100946 & 78249 & 28236 & 207428 \\
  02-May & 100704 & 79914 & 28710 & 209328 \\
  03-May & 100179 & 81654 & 28884 & 210717 \\
  04-May & 99980 & 82879 & 29079 & 211938 \\
  05-May & 98467 & 85231 & 29315 & 213013 \\
  06-May & 91528 & 93245 & 29684 & 214457 \\
  07-May & 89624 & 96276 & 29958 & 215858 \\
  08-May & 87961 & 99023 & 30201 & 217185 \\
  09-May & 84842 & 103031 & 30395 & 218268 \\
  10-May & 83324 & 105186 & 30560 & 219070 \\
  11-May & 82488 & 106587 & 30739 & 219814 \\
 12-May & 81266 & 109039 & 30911 & 221216 \\
 13-May & 78457 & 112541 & 31106 & 222104 \\
 14-May & 76440 & 115288 & 31368 & 223096 \\
 15-May & 72070 & 120205 & 31610 & 223885 \\
 16-May & 70187 & 122810 & 31763 & 224760 \\
 17-May & 68351 & 125176 & 31908 & 225435 \\
 18-May & 66553 & 127326 & 32007 & 225886 \\
 19-May & 65129 & 129401 & 32169 & 226699 \\
 20-May & 62752 & 132282 & 32330 & 227364 \\
 21-May & 60960 &  134560 & 32486 & 228006 \\
 22-May & 59322 & 136720 & 32616 & 229858 \\
 23-May & 57752 & 138840 & 32735 & 229327 \\
 24-May & 56594 & 140479 & 32785 & 229858 \\
 25-May & 55300 & 141981 & 32877 & 230158 \\
 26-May & 52942 & 144981 & 32955 & 230555 \\
 27-May & 50966 & 147101 & 33072 & 231139 \\
 28-May & 47986 &  150604 &33142 & 231732 \\
  \bottomrule
\end{tabular}
\caption{\footnotesize\textit{Situation in Italy on 15 May 2020}. Columns report the number of active people (currently infected by SARS-CoV-2), the number of recovered people, and the number of deceased people.}
\label{A1}
\end{table}

%% file: Data_BE.tex
\begin{table}[t]
\centering
\footnotesize
\begin{tabular}{lrrrr}
  \toprule
  Date & Active & Recovered & Deceased & Total cases\\
  \midrule
  
  29-Feb & 1 & 0 & 0 & 1 \\
  01-Mar & 1 & 0 & 0 & 1 \\
  02-Mar & 6 & 0 & 0 & 6 \\
  03-Mar & 13 & 0 & 0 & 13 \\
  04-Mar & 23 & 0 & 0 & 23 \\
  05-Mar & 50 & 0 & 0 & 50 \\
  06-Mar & 109 & 0 & 0 & 109 \\
  07-Mar & 169 & 0 & 0 & 169 \\
  08-Mar & 200 & 0 & 0 & 200 \\
  09-Mar & 239 & 0 & 0 & 239 \\
  10-Mar & 267 & 0 & 0 & 267 \\
  11-Mar & 311 & 0 & 3 & 314 \\
  12-Mar & 396 & 0 & 3 & 399 \\
  13-Mar & 556 & 0 & 3 & 599 \\
  14-Mar & 686 & 0 & 4 & 689 \\
  15-Mar & 881 & 1 & 4 & 886 \\
  16-Mar & 1052 & 1 & 5 & 1058 \\
  17-Mar & 1218 & 20 & 5 & 1243 \\
  18-Mar & 1441 & 31 & 14 & 1486 \\
  19-Mar & 1619 & 155 & 21 & 1795 \\
  20-Mar & 2016 & 204 & 37 & 2257 \\
  21-Mar & 2485 & 263 & 67 & 2815 \\
  22-Mar & 2986 & 340 & 75 & 3401 \\
  23-Mar & 3305 & 350 & 88 & 3743 \\
  24-Mar & 3737 & 410 & 122 & 4269 \\
  25-Mar & 4234 & 547 & 178 & 4937 \\
  26-Mar & 5340 & 675 & 220 & 6235 \\
  27-Mar & 6398 & 858 & 289& 7284 \\
  28-Mar & 7718 & 1063 & 353 & 9134 \\
  29-Mar & 9046 & 1359 & 431 & 10836\\
  30-Mar & 9859 & 1527 & 513 & 11899\\
  31-Mar & 10374 & 1696 & 705 & 12775\\
  01-Apr & 11004 & 2132 & 828 & 13964\\
  02-Apr & 11842 & 2495 & 1011 & 15348 \\
  03-Apr & 12755 & 2872 & 1143 & 16770\\
  04-Apr & 13901 & 3247 & 1283 & 18431\\
  05-Apr & 14493 & 3751 & 1447 & 19691\\
  06-Apr & 15196 & 3986 & 1632 & 20814\\
  07-Apr & 16002 & 4157 & 2035 & 22194\\
  08-Apr & 16482 & 4681 & 2240 & 23403\\
  09-Apr & 17296 & 5164 & 2523 & 24983\\
  10-Apr & 18080 & 5568 & 3019 & 26667\\
  11-Apr & 18686 & 5986 & 3346 & 28018\\
  12-Apr & 19584 & 6463 & 3600 & 29647\\
  13-Apr & 19979 & 6707 & 3903 & 30589\\
  14-Apr & 20094 & 6868 & 4157 & 31119\\
  15-Apr & 22025 & 7107 & 4440 & 33573\\
  \bottomrule
\end{tabular}
\caption{\footnotesize\textit{Situation in Belgium on 15 May 2020}. Columns report the number of active people (currently infected by SARS-CoV-2), the number of recovered people, and the number of deceased people.}
\label{B}
\end{table}

%% file: Data_BE_II.tex
\begin{table}[t]
\centering
\footnotesize
\begin{tabular}{lrrrr}
  \toprule
  Date & Active & Recovered & Deceased & Total cases\\
  \midrule
  
  16-Apr & 22390 & 7562 & 4857 & 34809 \\
  17-Apr  & 23014 & 7961 & 5163 & 36138 \\
  18-Apr  & 23346 & 8384 & 5453 & 37183 \\
  19-Apr & 24056 & 8757 & 5683 & 38496 \\
  20-Apr  & 25260 & 8895 & 5828 & 39983 \\
  21-Apr  & 25296 & 9002 & 5998 & 40956 \\
  22-Apr  & 26194 & 9433 & 6262 & 41889 \\
  23-Apr  & 26507 & 9800 & 6490 & 42797 \\
  24-Apr & 27492 & 10122 & 6679 & 44293 \\
  25-Apr & 27991 & 10417 & 6917 & 45325 \\
  26-Apr & 28255 & 10785 & 7094 & 46134 \\
  27-Apr  & 28602 & 10878 & 7207 & 46687 \\
  28-Apr  & 29060 & 10943 & 7331 & 47334 \\
  29-Apr  & 29075 & 11283 & 7501 & 47859 \\
  30-Apr  & 29349 & 11576 & 7594 & 48519 \\
  01-May & 29437 & 11892 & 7703 & 49032 \\
  02-May & 29541 & 12211 & 7765 & 49517 \\
  03-May& 29753 & 12309 & 7844 & 49906 \\
  04-May & 29965 & 12378 & 7924 & 50267 \\
  05-May & 30052 & 12441 & 8016 & 50509 \\
  06-May & 29711 & 12731 & 8339 & 50781 \\
  07-May & 30025 & 12980 & 8415 & 51420 \\
  08-May & 30289 & 13201& 8521 & 52011 \\
  09-May & 30604 & 13411 & 8581 & 52596 \\
10-May & 30783 & 13642 & 8656 & 53081 \\
11-May& 31045 & 13697 & 8707 & 53449 \\
12-May & 31286 & 13732 & 8761 & 53779 \\
13-May & 31201 & 13937 & 8843 & 53981 \\
14-May& 31274 & 14111 & 8903 & 54288 \\
15-May & 31384 & 14301 & 8959 & 54644\\
16-May & 31524 & 14460 & 9005 & 54986\\
17-May & 31598 & 14630 & 9052 & 55280\\
18-May & 31822 & 14657 & 9080 & 55559\\
19-May & 31996 & 14687 & 9108 & 55791\\
20-May & 31986 & 14847 & 9150 & 55983\\
21-May & 32061 & 14988 & 9186 & 56235\\
22-May & 32176 & 15123 & 9212 & 56810\\
23-May & 32418 & 15155 & 9237 & 56810\\
24-May & 32540 & 15272 & 9280 & 57092\\
25-May & 32733 & 15297 & 9312 & 57342\\
26-May & 32801 & 15320 & 9334 & 57455\\
27-May & 32763 & 15465 & 9364 & 57592\\
28-May & 32889 & 15572 & 9388 & 57849\\
  \bottomrule
\end{tabular}
\caption{\footnotesize\textit{Situation in Belgium on 15 May 2020}. Columns report the number of active people (currently infected by SARS-CoV-2), the number of recovered people, and the number of deceased people.}
\label{B1}
\end{table}

%% file: Data_LUX.tex
\begin{table}[t]
\centering
\footnotesize
\begin{tabular}{lrrrr}
  \toprule
  Date & Active & Recovered & Deceased & Total cases\\
  \midrule
  29-Feb & 1 & 0 & 0 & 1 \\
  01-Mar & 1 & 0 & 0 & 1 \\
  02-Mar & 1 & 0 & 0 & 1 \\
  03-Mar & 1 & 0 & 0 & 1 \\
  04-Mar & 1 & 0 & 0 & 1 \\
  05-Mar & 2 & 0 & 0 & 2 \\
  06-Mar & 3 & 0 & 0 & 4 \\
  07-Mar & 4 & 0 & 0 & 4 \\
  08-Mar & 5 & 0 & 0 & 5 \\
  09-Mar & 5 & 0 & 0 & 5 \\
  10-Mar & 7 & 0 & 0 & 7 \\
  11-Mar & 7 & 0 & 0 & 7 \\
  12-Mar & 26 & 0 & 0 & 26 \\
  13-Mar & 33 & 0 & 1 & 34 \\
  14-Mar & 50 & 0 & 1 & 51 \\
  15-Mar & 76 & 0 & 1 & 77 \\
  16-Mar & 80 & 0 & 1 & 81 \\
  17-Mar & 139 & 0 & 1 & 140 \\
  18-Mar & 201 & 0 & 2 & 203 \\
  19-Mar & 325 & 6 & 4 & 335 \\
  20-Mar & 474 & 6 & 4 & 484 \\
  21-Mar & 656 & 6 & 8 & 670 \\
  22-Mar & 780 & 10 & 8 & 798 \\
  23-Mar$\star$ & 857 & 10 & 8 & 875 \\
  24-Mar & 1081 & 10 & 8 & 1099 \\
  25-Mar & 1315 & 10 & 8 & 1333 \\
  26-Mar & 1434 & 10 & 9 & 1453 \\
  27-Mar$\star\star$ & 1550 & 40 & 15 & 1605 \\
  28-Mar & 1773 & 40 & 18 & 1831 \\
  29-Mar & 1889 & 40 & 21 & 1950 \\
30-Mar & 1926 & 40 & 22 & 1988 \\
  31-Mar & 2115 & 40 & 23 & 2178 \\
  01-Apr$\star\star$ & 2250 & 40 & 29 & 2319 \\
  02-Apr & 2417 & 40 & 30 & 2487 \\
  03-Apr & 2081 & 500 & 31 & 2612 \\
  04-Apr & 2198 & 500 & 31 & 2729 \\
  05-Apr & 2268 & 500 & 36 & 2804 \\
  06-Apr & 2302 & 500 & 41 & 2843 \\
  07-Apr & 2426 & 500 & 44 & 2970 \\
  08-Apr & 2488 & 500 & 46 & 3034 \\
   09-Apr & 2563& 500 & 52 & 3115 \\
   10-Apr & 2669 & 500 & 54 & 3223 \\
   11-Apr & 2708 & 500 & 62 & 3270 \\
   12-Apr & 2715 & 500 & 66 & 3281 \\
   13-Apr & 2725 & 500 & 67 & 3292 \\
   14-Apr & 2740 & 500 & 67 & 3307 \\
   15-Apr & 2778 & 520 & 69 & 3373 \\
  \bottomrule
\end{tabular}

\noindent 
$\star$ Attention on the 23rd the figures instead of being given at 9 am are given at 5 pm. 

$\star\star$ Including 1 evacuated from France.

\caption{\footnotesize\textit{Situation in Luxembourg on 15 May 2020}. Columns provide the number of active people (currently infected by SARS-CoV-2), the number of recovered people, and the number of deceased people.} 
\label{C}
\end{table}

%% file: Data_LUX_II.tex
\begin{table}[t]
\centering
\footnotesize
\begin{tabular}{lrrrr}
  \toprule
  Date & Active & Recovered & Deceased & Total cases\\
  \midrule
  16-Apr & 1008 & 2368 & 68 & 3444 \\
  17-Apr  & 919 & 2489 & 72 & 3480 \\
  18-Apr  & 858 & 2607 & 72 & 3537 \\
  19-Apr  & 799 & 2678 & 73 & 3550 \\
  20-Apr  & 759 & 2724 & 75 & 3558 \\
  21-Apr  & 735 & 2805 & 78 & 3618 \\
  22-Apr  & 689 & 2885 & 80 & 3654 \\
  23-Apr  & 619 & 2963 & 83 & 3665 \\
  24-Apr  & 582 & 3028 & 85 & 3695 \\
  25-Apr  & 539 & 3087 & 85 & 3711 \\
  26-Apr  & 532 & 3103 & 88 & 3723 \\
  27-Apr  & 520 & 3121 & 88 & 3729 \\
  28-Apr  & 531 & 3121 & 89 & 3741 \\
  29-Apr  & 546 & 3134 & 89 & 3769 \\
  30-Apr  & 481 & 3213 & 90 & 3784 \\
  01-May & 497 & 3213 & 92 & 3802 \\
  02-May & 402 & 3318 & 92 & 3812 \\
  03-May & 349 & 3379 & 96 & 3824 \\
  04-May & 327 & 3405 & 96 & 3828 \\
  05-May & 332 & 3412 & 96 & 3840 \\
  06-May & 301 & 3452 & 98 & 3851 \\
  07-May & 254 & 3505 & 100 & 3859 \\
  08-May & 245 & 3526 & 100 & 3871 \\
  09-May & 226 & 3550 & 101 & 3877 \\
  10-May & 199 & 3586 & 101 & 3886 \\
  11-May & 185 & 3602 & 101 & 3888 \\
  12-May & 182 & 3610 & 102 & 3894 \\
  13-May & 172 & 3629 & 103 & 3904 \\
  14-May & 147 & 3665 & 103 & 3915 \\
 15-May & 137 & 3682 & 104 & 3923 \\
 16-May & 127 & 3699 & 104 & 3930 \\
 17-May & 136 & 3702 & 107 & 3945 \\
 18-May & 125 & 3715 & 107 & 3947 \\
 19-May & 131 & 3718 & 109 & 3958 \\
 20-May & 134 & 3728 & 109 & 3971 \\
 21-May & 130 & 3741 & 109 & 3980 \\
 22-May & 124 & 3748 & 109 & 3981 \\
 23-May & 123 & 3758 & 109 & 3990 \\
 24-May & 115 & 3767 & 110 & 3992 \\
 25-May & 102 & 3781 & 110 & 3993 \\
 26-May & 102 & 3783 & 110 & 3995 \\
 27-May & 100 & 3791 & 110 & 4001 \\
 28-May & 95 & 3803 & 110 & 4008 \\
  \bottomrule
\end{tabular}

\caption{\footnotesize\textit{Situation in Luxembourg on 15 May 2020}. Columns provide the number of active people (currently infected by SARS-CoV-2), the number of recovered people, and the number of deceased people.} 
\label{C1}
\end{table}